\documentclass[12pt,english,oneside,BCOR=12mm,abstract=on]{scrartcl}
\usepackage[english]{babel}
\usepackage{graphicx}			
\usepackage{amsmath}			
\usepackage{bbm}				
\usepackage{amsfonts} 			
\usepackage{yfonts}				
\usepackage{cjhebrew}			
\usepackage[thref,amsmath,thmmarks]{ntheorem}			
\usepackage{amsrefs}			
\usepackage{enumerate}			
\usepackage{longtable}			
\usepackage{array}				
\usepackage{accents}      

\newcommand{\abs}[1]{\lvert#1\rvert}	
\newcommand{\norm}[1]{\lVert#1\rVert}	
\newcommand{\set}[2]{\left\lbrace #1 \middle\vert #2 \right\rbrace}
	
\newcommand{\Lieb}[2]{\left[ #1 , #2 \right]}	
\newcommand{\inp}[2]{\left\langle #1 \text{,} #2 \right\rangle} 

\newcommand{\EXP}[1]{\mathbb E \left( #1 \right)}		
\newcommand{\partd}[2]{\frac{\partial #1}{\partial #2}}	
\renewcommand{\d}{\text{\textnormal{d}}}				
\renewcommand{\abs}[1]{\left\lvert #1 \right\rvert}

\newcommand{\tp}{\otimes}				
\newcommand{\cross}{\times}				
\newcommand{\R}{\mathbb{R}}				
\newcommand{\C}{\mathbb{C}}				
\newcommand{\Z}{\mathbb{Z}}				
\newcommand{\N}{\mathbb{N}}				
\let\ndoti\i							
\renewcommand{\i}{\mathfrak{i}}			
\DeclareMathSymbol{\varnothing}{\mathord}{AMSb}{"3F}
\renewcommand{\emptyset}{\varnothing}	
\DeclareMathSymbol{\upharpoonright} {\mathrel}{AMSa}{"16}

\DeclareMathOperator{\dom}{dom}
\DeclareMathOperator{\curl}{curl}
\DeclareMathOperator{\grad}{grad}
\let\div\relax
\DeclareMathOperator{\div}{div}

\let\Re\relax
\DeclareMathOperator{\Re}{Re}
\let\Im\relax
\DeclareMathOperator{\Im}{Im}

\DeclareMathOperator{\End}{End}				
\DeclareMathOperator{\GenL}{GL}    			
\DeclareMathOperator{\Frameb}{Fr}    			
\DeclareMathOperator{\bigo}{\mathcal O}	
\DeclareMathOperator{\dist}{dist}   		
\DeclareMathSymbol{\square}
{\mathord}{AMSa}{"03}
\DeclareMathSymbol{\blacksquare} {\mathord}{AMSa}{"04}
\newcommand{\evat}[1]{_{\upharpoonright_{#1}}}
\newcommand{\Evat}[2]{\left. #1 \right\rvert_{#2}}
\renewcommand{\qedsymbol}{
								$\blacksquare$
							}
\newcommand{\oendmark}{	
						$\diamondsuit$
						}

\theoremstyle{break}
\theoremheaderfont{\normalfont\bfseries}
\theorembodyfont{\normalfont
				}
\theoremseparator{}
\theoremnumbering{arabic}
\theoremsymbol{\oendmark}
\newtheorem{Proposition}{Proposition}[section]	
\newtheorem{Theorem}[Proposition]{Theorem}
\newtheorem{Lemma}[Proposition]{Lemma}
\newtheorem{Corollary}[Proposition]{Corollary}

\theoremsymbol{\oendmark}
\newtheorem{Definition}[Proposition]{Definition}
\newtheorem{Principle}{Principle}

\newtheorem{Postulate}{Postulate}

\newtheorem{Remark}[Proposition]{Remark}

\newtheorem{Question}{Question}

\theoremstyle{nonumberplain}
\theoremseparator{:}
\theoremsymbol{\qedsymbol} 
\newtheorem{Proof}{Proof}
\numberwithin{equation}{section}

\newcolumntype{M}[1]{>{\centering\arraybackslash}m{#1}} 

\begin{document}     

\title{  
		The Madelung Picture as a Foundation of 
		Geometric Quantum Theory
		}

\author{	Maik Reddiger	\\
			\texttt{maik.reddiger@campus.tu-berlin.de
			}}
\date{\today}
\dedication{\footnotesize{\emph{Keywords:} Geometric Quantization \, - \, Interpretation of Quantum Mechanics \, - \, Geometric Quantum Theory \, - \, Madelung Equations \, - \, Classical Limit}}

\maketitle 

\begin{abstract} 
	\noindent
	Despite its age, quantum theory still suffers from serious conceptual 
	difficulties. To create clarity, mathematical physicists have been
	attempting to formulate quantum theory geometrically and to 
	find a rigorous method of quantization, but this has not resolved the 
	problem.\\
	In this article we argue that a quantum theory recursing 
	to quantization algorithms is necessarily incomplete. To provide 
	an alternative approach, we show that the Schr\"odinger equation is 
	a consequence of three partial 
	differential equations governing the time evolution of a given
	probability density. These equations, discovered by E. Madelung, 
	naturally ground the Schr\"odinger theory in Newtonian mechanics and 
	Kolmogorovian probability theory. A variety of far-reaching 
	consequences for the projection postulate, the 
	correspondence principle, the measurement problem, 
	the uncertainty principle, and the modeling of particle creation 
	and annihilation are immediate. We also give a speculative 
	interpretation of the equations following Bohm, Vigier and Tsekov, 
	by claiming that quantum mechanical behavior is possibly caused by 
	gravitational background noise.
\end{abstract}
\noindent

\newpage 

\tableofcontents      

\section{Introductory Discussion}  
\label{sec:intro} 
\subsection{Critique of Quantization and a new Methodology} 

The idea of quantization was first put forward by Dirac 
\cite{Dirac} in 1925 in an attempt to extend Heisenberg's theory of 
matrix mechanics \cite{Heisenberg}. He based the concept on a 
formal analogy between the Hamilton and the Heisenberg equation and 
on the principle of correspondence, namely that a quantum 
theoretical model should yield a ``classical'' one in some limit. 
This analogy motivated Dirac to develop a scheme that constructs 
one or more quantum analogues of a given ``classical system'' 
formulated in the language of Hamiltonian 
mechanics.\footnote{``In a recent paper Heisenberg puts forward a 
new theory, which suggests that it is not the equations of 
classical mechanics that are in any way at fault, but that 
the mathematical operations by which physical results are deduced 
from them require modification." \cite{Dirac}} 
When it was discovered that Dirac's scheme, nowadays known as 
canonical quantization, was ill-defined
(see \cites{Groenewold,Hove} for the original works by Groenewold and 
van Hove, also \cite{Abraham}*{\S 5.4}, in particular	
\cite{Abraham}*{Thm. 5.4.9}), physicists and mathematicians attempted 
to develop a more sophisticated machinery rather than questioning 
the ansatz. The result has been a variety of quantization algorithms, 
one of which is particularly noteworthy: Geometric quantization 
(cf. \cites{Woodhouse,Enriquez} for an introduction). 
\par 
In his seminal paper, Segal \cite{Segal} expressed the need to employ 
the language of differential geometry in quantum theory. He 
understood that determining the relevant differential-geometric 
structures, spaces and their relation to the fundamental equations 
of quantum theory creates the mathematical coherence necessary 
to adequately address foundational issues in the subject. By merging 
this ansatz with Kirillov's work in representation theory \cite{Kirillov}, 
Segal, Kostant \cite{Kostant} and Souriau \cite{Souriau} were able 
to construct the algorithm of geometric quantization. However, rather 
than elaborating on the relation between quantum and classical 
mechanics, geometric quantization unearthed a large amount of 
geometric structures \cite{Hall}*{\S 23.2}, introduced 
in an ad hoc manner.
\par 
It is tempting to blame this state of affairs on the inadequacy of 
the geometric ansatz or the theory, but instead we invite the reader 
to take a step back. What is the reason for the construction of 
a quantization algorithm? Why do we quantize? Certainly, quantum mechanics 
should agree with Newtonian mechanics in some approximation, 
where the latter is known to 
accord with experiment, but is it reasonable to assume the existence of 
an algorithm that constructs the new theory out of the old one? 
\par 
These questions are of philosophical nature and it is useful to 
address them within the historical context. Clearly, the step 
from Newtonian mechanics to quantum mechanics was a scientific 
revolution, which is why we find the work of the philosopher and 
physicist Thomas Kuhn \cite{Kuhn} of relevance to our discussion. Kuhn 
is known for his book ``The Structure of Scientific Revolutions'' 
\cite{Kuhn}, in which he analyzed the steps of scientific progress in 
the natural sciences. For a summary see \cite{Pajares}. 
\par
Kuhn argues that, as a field of science develops, a paradigm 
is eventually formed through which all empirical data is 
interpreted. As, however, the empirical evidence becomes 
increasingly incompatible with the paradigm, it is modified in 
an ad hoc manner in order to allow for progress in the field. 
Ultimately, this creates a crisis,  
as attempts to 
account for the evidence become increasingly ad hoc, unmanageably 
elaborate and ultimately contradictory. Unless a new paradigm is 
presented and withstands experimental and theoretical scrutiny, the 
crisis persists and deepens, because of the internal and 
external inconsistencies of the current paradigm. 
\par
This process can be directly observed in the history of quantum theory. 
When Newtonian mechanics was faced with the problem of describing 
the atomic spectra and the stability of the atom in the beginning 
of the twentieth century \cite{Bohr1}, it was ad hoc modified by 
adding the Bohr-Sommerfeld quantization condition 
\cites{Bohr1, Sommerfeld}, despite its known inconsistency with 
then accepted principles of physics \cites{Bohr2,Heisenberg1}. 
This ad hoc modification of Newtonian mechanics continued 
with Werner Heisenberg's \cite{Heisenberg} and Erwin 
Schr\"odinger's \cite{Schroedinger1} postulation of their 
fundamental equations of quantum mechanics, two descriptions 
later shown to be formally equivalent by von Neumann in his 
constitutive work \cite{Neumann}. Schr\"odinger's and Heisenberg's  
description can be viewed as an ad hoc modification, because 
their equations are formulated on a Newtonian spacetime and 
intended to replace Newton's second law without being based on 
postulated principles of nature. With his quantization algorithm 
\cite{Dirac}, Dirac supplied a convenient way to pass 
from the mathematical description of a physical system in 
Newtonian mechanics to the then incomplete, new theory. In 
accordance with Kuhn's description, it was a pragmatic, ad hoc 
step, not one rooted in deep philosophical reflection. Nonetheless, 
the concept of quantization is ingrained in quantum theory as of 
today \cite{Weinberg0}, while the as of now futile search for 
unity in physics has become increasingly ad hoc and 
elaborate \cite{Woit}*{\S 19}.
\par
We are thus reminded of our historical position and the original 
intention behind quantization: We would like to be able to 
mathematically describe microscopic phenomena, having at hand 
neither the fundamental equations describing those phenomena nor 
a proper understanding of the physical principles involved allowing 
us to derive such equations. That is, what we lack with respect 
to our knowledge of microscopic phenomena is, in Kuhn's words, 
a paradigm. Rather than having a set of principles of nature, 
which we use to intuitively understand and derive the fundamental 
laws of quantum theory, we physicists assume the validity of the 
old theory, namely Newtonian mechanics or special relativity in 
its Hamiltonian formulation, only to apply an ad hoc algorithm 
to obtain laws we have inadequately understood. This is why 
the concept of quantization itself is objectionable. 
\par
Indeed, even if a mathematically well defined quantization scheme 
existed, it would remain an ad hoc procedure and one would still need 
additional knowledge which quantized systems are physical 
(cf. \cite{Waldmann}*{\S 5.1.2} for a discussion of this in German). 
From a theory builder's perspective, it would then be more favorable 
to simply use the quantized, physically correct models as a 
theoretical basis and deduce the classical models out of these, 
rather than formulating the theory in the reverse way. Hence 
quantization can be viewed as a procedure invented to systematically guess 
quantum-theoretical models. This is done with the implicit expectation 
of shedding some light on the conceptual and mathematical 
problems of quantum theory, so that one day a theory can be deduced 
from first principles. Thus a quantum theory, which is constructed from 
a quantization scheme, must necessarily be incomplete. More precisely, 
it has not been formulated as a closed entity, since for its 
formulation it requires the theory it attempts to replace and 
which it potentially contradicts. 
\par
As a result of this development, quantum mechanics and thus quantum 
theory as a whole has not been able to pass beyond its status as an 
ad hoc modification of Newtonian 
mechanics and relativity to date. For a recapitulation of the history 
of quantum theory illustrating this point, see e.g.
the article by Heisenberg \cite{Heisenberg1}.   
\par   
Fortunately, our criticism does not apply to the theory of 
relativity, which to our knowledge provides an accurate description 
of phenomena \cite{Will}, at least in the macroscopic realm. 
As the principles of relativity theory are known (cf. 
\cite{Kriele}*{p. XVII}), the ridiculousness of ``relativizing'' 
Newtonian mechanics is obvious. Indeed, in the theory of 
relativity physics still finds a working paradigm. 
\par 
Rejecting quantization neither leads to a rejection of quantum 
theory itself, nor does it imply that previous attempts to put 
quantum theory into a geometric language were futile. If we 
reject quantization, we are forced to view quantum theory as 
incomplete and phenomenological, which raises the question of what 
the underlying physical principles and observables are. Considering 
that the theory of relativity is mainly a theory of spacetime 
geometry, asking, as Segal did, for the primary geometric and 
physical quantities in quantum theory offers a promising and 
natural approach to this question. 
\par
Therefore, we reason that we theorists should look at the equations 
of quantum theory with strong empirical support and use these to 
construct a mathematically consistent, probabilistic, geometric 
theory, tied to fundamental physical principles as closely 
as possible. But how is this to be approached?

\subsection{The Madelung Equations as a Geometric Ansatz}

In the year 1926, the same year Schr\"odinger published his 
famous articles \cites{Schroedinger1,Schroedinger2,Schroedinger3}, 
the German physicist Erwin Madelung reformulated the 
Schr\"odinger equation into a set of real, non-linear partial 
differential equations \cite{Madelung} with strong resemblance to 
the Euler equations \cite{Chorin}*{\S 1.1} found in hydrodynamics. 
The so-called \emph{Madelung equations} are%
\footnote{Here we use the usual notation for vector calculus 
on $\R^3$ with standard metric $\delta$.} 
\begin{gather}
	m \dot{\vec{X}} = \vec F + \frac{\hbar^2}{2m} 
	\nabla \frac{\Delta \sqrt{\rho}}{\sqrt{\rho}}  \, ,\\
	\nabla \cross \vec X = 0 \, ,\\
	\frac{\partial \rho}{\partial t}	+ 
	\nabla \cdot \left( \rho \, \vec X \right) = 0 \, ,  
\end{gather} 
where $m$ is the mass of the particle, $X = \partial/ \partial t + 
\vec X$ is a real vector field, called the \emph{drift (velocity) 
field}, $\rho$ is the \emph{probability density} 
(by an abuse of terminology), 
$\vec F$ the external force and $\dot{\vec{X}}$ denotes the 
so-called material derivative (cf. \cite{Chorin}*{p. 4}) of $X$ 
along itself. 
Madelung already believed%
\footnote{``Es besteht somit Aussicht auf dieser Basis die 
Quantentheorie der Atome zu erledigen.'' 
\cite{Madelung}*{p. 326}; translation by author: 
``There is hence a prospect to complete the quantum 
theory of atoms on this basis.''}
 that these equations could serve as a foundation of quantum 
theory. He reached this conclusion, because the equations exhibit 
a strong link between quantum mechanics and Newtonian continuum 
mechanics \cite{Madelung}. Thus Madelung used these equations 
to interpret quantum behavior by exploiting the analogy to the 
Euler equations. At this point in history, it was not clear 
how to interpret the wave function as the Born rule 
and the ensemble interpretation had just recently emerged \cite{Born}. 
Madelung's misinterpretation of quantum mechanics may 
perhaps be the reason why it took almost 25 years for his approach 
to become popular again, when Bohm employed the Madelung equations 
to develop what is now known as Bohmian mechanics 
\cites{Bohm1,Bohm2}. Nonetheless a clear distinction should 
be drawn \cite{Tsekov2} between the Madelung equations and the 
Bohmian theory \cites{Bohm1,Bohm2}. 
Despite the popularity of Bohm's approach, a discussion of 
the Madelung equations on their own 
\cites{Holland1,Janossy1,Janossy2,Janossy3,Takabayasi,Wallstrom}
seems less common.  
\par 
Today, the importance of the Madelung equations lies in the fact 
that they naturally generalize the Schr\"odinger equation and in 
doing so expose the sought-after geometric structures of quantum 
theory and its classical limit. As a byproduct, one obtains a 
natural answer to the question why complex numbers arise in 
quantum mechanics.  
The Madelung equations, by their virtue of being formulated 
in the language of Newtonian mechanics, make it possible to 
construct a wide 
class of quantum theories by making the same 
coordinate-independent modifications found in Newtonian mechanics, 
without any need to construct a quantization algorithm as, for 
example, in geometric \cites{Woodhouse,Enriquez} and 
deformation quantization \cite{Waldmann}. This greatly 
simplifies the construction of new quantum theories and 
therefore makes the Madelung equations the natural foundation of 
quantum mechanics and the natural ansatz for any attempts 
of interpreting quantum mechanics. 
\par
For some of these modifications it is not possible
to construct a Schr\"odinger equation and for others the 
Schr\"odinger equation becomes non-linear, which suggests that 
there exist quantum-mechanical models that cannot be formulated 
in the language of linear operators acting on a vector 
space of 
functions. From a conceptual point of view, this might prove to be 
a necessity to remove the mathematical and conceptual problems 
that plague relativistic quantum theory today or at least 
expose the origins of these problems. 
In fact, the Madelung equations admit a straight-forward 
(general-)relativistic 
generalization leading to the Klein-Gordon equation, which is, 
however, not discussed here and arguably unphysical.%
\footnote{We believe that the lack of physicality is 
a consequence of neglecting spin in the Schr\"odinger theory. We refer 
to \citelist{\cite{Gurtler1} \cite{Hestenes1}} 
for an elaboration on this point of view.
\label{ftn:KG}} 
The Madelung equations and 
their modifications are henceforth particularly suited for 
studying quantum theory from the differential-geometric perspective. 
We thus believe that they will take a central role both in 
the future construction of an internally consistent, 
\emph{geometric quantum theory} as well as the realist 
understanding of microscopic phenomena.

\subsection{Outline and Conventions}

In this article we formalize the Madelung picture of quantum mechanics 
and thus provide a rigorous framework for further development. A 
first step is made by postulating a modification intended to 
model particle creation and annihilation. In addition, we give a 
possible interpretation of quantum mechanics that is an extension 
of the stochastic interpretation developed by Tsekov \cite{Tsekov1}, 
which in turn originated in ideas from Bohm and Vigier \cite{Bohm3} 
in the 1950s. 
\par
Our article is organized as follows: We first construct a spacetime 
model on which to formulate the Madelung equations using 
relativistic considerations. In section \ref{sec:equivalence} on 
page \pageref{sec:equivalence}, we further motivate the need for 
the Madelung equations in the formulation of quantum mechanics 
and then give a theorem stating the equivalence of the Madelung 
equations and the Schr\"odinger equation, if the force is irrotational 
and a certain topological condition is satisfied. We also 
address concerns raised in the literature \citelist{\cite{Takabayasi} 
\cite{Wallstrom0} \cite{Wallstrom} \cite{Holland}*{\S 3.2.2}} regarding 
this point. 
We introduce some terminology and proceed with a basic, 
mathematical discussion. In section 
\ref{sec:operator} on page \pageref{sec:operator}, we discuss the operator
formalism in the Schr\"odinger picture and its relation to the 
Madelung equations. We proceed by giving a formal interpretation of 
the Madelung equations in section \ref{sec:interpretation1} 
on page \pageref{sec:interpretation1} and then speculate 
in section \ref{sec:interpretation2} on page 
\pageref{sec:interpretation2} that quantum mechanical behavior 
originates in noise created by random irregularities in 
spacetime curvature, that is random, small-amplitude gravitational 
waves. How the 
violation of Bell's inequality can be achieved in this 
stochastic interpretation is also discussed. In section 
\ref{sec:annihilation} on page \pageref{sec:annihilation}, 
we propose a modification of the Madelung equations, intended 
to model particle creation and 
annihilation, and show how this in general leads to a non-linearity 
in the Schr\"odinger equation. We conclude this article on page 
\pageref{sec:conclusion} with a brief review of our results 
including a table and an overview of some open problems. 
\par
Some prior remarks: To fully understand this article, an 
elementary knowledge of Riemannian geometry, relativity and 
quantum mechanics is required. We refer to 
\cite{Rudolph}*{Chap. 1-4}, \cites{Carroll,Wald} and 
\cite{Ballentine}, respectively. The mathematical formalism of the 
article is, however, not intended to deter anyone from reading it 
and should not be a hindrance to understanding the physics we 
discuss, which is not merely of relevance to mathematical physicists. 
For the sake of clarification, we have attempted to provide some 
intuitive insight along the lines of the argument. Less 
mathematically versed readers should skip the proofs and the 
more technical arguments while being aware that precise 
mathematical arguments are required, as intuition fails easily 
in a subject this far away from everyday experience. 
Moreover, we stress that section \ref{sec:interpretation2} should 
be considered fully separate from the rest of the article. At this 
point the stochastic interpretation, however well motivated, 
is speculation, but this does not invalidate the rest of the argument. 
\par
On a technical note, we usually assume that all mappings and 
manifolds are smooth. This assumption can be considerably 
relaxed in most cases, but this would lead to additional, 
currently unnecessary technicalities. Our notation mostly 
originates from \cite{Rudolph}, but is quite standard in physics 
or differential geometry. For example, $\varphi_*$ is the 
pushforward and $\varphi^*$ the pullback of the smooth map 
$\varphi$, $\cdot$ is tensor contraction of adjacent entries or the 
Euclidean inner product, 
$\d$ the Cartan derivative, $\varepsilon$ the Levi-Civita symbol, 
$\Lieb{.}{.}$ the Lie bracket (of vector fields), 
$\mathfrak X \left( \mathcal Q \right)$ denotes the space of smooth 
vector fields and $\Omega^k \left( \mathcal Q \right)$ the space of 
smooth $k$-forms on the smooth manifold $\mathcal Q$, respectively. 
We use the Einstein summation convention and, where relativistic 
arguments are used, the metric signature is $(+---)$, which 
gives tangent vectors of observers positive ``norm''.   
Definitions are indicated by \emph{italics}. 

\section{Construction of Newtonian Spacetime}
\label{sec:Newton}

In order to be able to construct a rigorous proof of the 
equivalence of the Schr\"odinger and Madelung equations, we 
first construct a spacetime model suitable for our purposes. 
For a discussion on prerelativistic spacetimes see e.g. 
\cite{Kriele}*{\S1.1 to \S1.3} and \cite{ArnoldV}*{Chap. 1}.   
\par
To describe the motion of a point mass of mass 
$m \in \R_+ = \left(0 , \infty \right)$ in Newtonian physics, 
we consider an open subset $\mathcal Q$ of $\R^4$, which has a 
canonical topology and smooth structure.
The need to restrict oneself to open subsets of $\R^4$ arises, 
for instance,  from the fact that it is common for forces in 
Newtonian physics to diverge at the point where the source is 
located. We exclude such points from the manifold. For similar 
reasons we also allow non-connected subsets. 
\par
To be able to measure spatial distances within the Newtonian 
ontology, one intuitively needs a degenerate, Euclidean metric. 
However, this construction should obey the principle 
of Galilean relativity (cf. \cite{Kriele}*{Postulate 1.3.1}). 
\begin{Principle}[Galilean relativity]
	\label{Prin:Galileo}
	For any two non-accelerating observers that move relative to 
	each other with constant velocity all mechanical processes are 
	the same. 
\end{Principle}
Therefore, if we formulate physical laws coordinate-independently 
with some (degenerate) metric $\delta$ and attribute to it a 
physical reality, then all observers should measure the same 
distances. However, in physical terms, whether one travels 
some distance at constant velocity or is standing still, fully 
depends on the observer, hence the coordinate system chosen 
to describe the system. This is a deep problem within the conceptual 
framework of Newtonian mechanics. One way to circumvent this, is 
to prevent the measurement of distances for different times. 
For a mathematical treatment of such Neo-Newtonian or, better 
to say, Galilean spacetimes see \cite{ArnoldV}*{Chap. 1}. 
A less complicated and physically more satisfying approach is to 
consider a Newtonian spacetime as a limiting case of a 
special-relativistic one. More precisely, a Newtonian spacetime is 
an approximative spacetime model appropriate for mechanical 
systems involving only small velocities relative to an inertial 
frame of reference and relative 
to the speed of light, not involving the modeling of light 
itself and with negligible spacetime curvature. In this 
relativistic ontology, the above conceptual problem does not occur, 
as the notion of spatial and temporal distance is made 
observer-dependent, which is necessary due to the phenomenon 
of time dilation and length contraction. 
\par 
As quantum mechanics is formulated in a Newtonian/Galilean 
spacetime, it is consequently necessary to view it as a theory 
in the so-called \emph{Newtonian limit}. This limit is naively 
 defined by  neglecting terms of the order $\bigo \left( \left( 
\abs{\vec v} / c \right)^2 \right)$ in equations involving only 
physically measurable quantities, where $\abs{\vec v}$ is the 
speed corresponding to the velocity $\vec v$ of any mass point 
relative to the inertial frame and $c$ is the speed of light 
(in vacuum). 
Obviously, this is not a rigorous definition, but 
this naive approach suffices for our purposes here.
We will 
give a more thorough discussion of the Newtonian limit 
in a future work \cite{ReddigerB0}. 
Also note that $\abs{\vec v} / c$ is dimensionless and hence the 
Newtonian limit is independent of the chosen system of units.
\par 
Our reasoning directly leads us to the definition of 
Newtonian spacetime. 
\begin{Definition}[Newtonian spacetime]
	\label{Def:Nst}
\begin{subequations}
	A \emph{Newtonian spacetime} is a tuple 
	$\left( \mathcal Q , \d \tau, \delta, 
	\mathcal O \right)$, where 
	\begin{enumerate}[i)]
		\item $\mathcal Q$ is an open subset of 
				$\R^4$ equipped with the standard topology and
				smooth structure, 
		\item	\label{Item:metrics}
				the \emph{time form} $\d \tau$ is an exact, 	
				non-vanishing $1$-form and the \emph{spatial 
				metric} $\delta$ is a symmetric, 
				non-vanishing, covariant $2$-tensor field,
				such that there exist coordinates $x= \left(t, 
				x^1, x^2, x^3 \right) \equiv \left( t, 
				\vec x \right)$ on $\mathcal Q$ 
				with 
				\begin{equation}
					\d \tau = \d t \, , \quad 
					\delta = \delta_{ij} \,  \d x^i \tp \d x^j
					=
					\begin{pmatrix}
						0 & & & \\
						& 1 & & \\
						& & 1 & \\
						& & & 1
					\end{pmatrix}
					\label{eq:Eulercond}
				\end{equation}
				for $i,j \in \lbrace 1,2, 3 \rbrace$. 
		\item	\label{Item:Norient}
				the \emph{Newtonian orientation} $\mathcal O$ is a 
				(smooth) 
				$\GenL^+\left( \R^3 \right)$-reduction
				of the frame bundle $\Frameb \left(T \mathcal Q
				\right)$ defined as follows (see e.g. 
				\citelist{\cite{Rudolph}*{\S 6.1} 
				\cite{Poor}*{Def. 9.6}} for 
				definitions).
				Consider the Lie group 
					\begin{equation}
						\GenL^+\left( \R^3 \right) := 
						\set {A \in \End \left( \R^3 \right)}{
						\det A > 0} \, ,
					\end{equation}	 
				the vector field $B$, defined by 
				\begin{align}
					\delta \left( B, B\right) &= 0 \, , 
					\label{eq:Bspace} \\
	 				\d \tau \cdot B &= 1  \, ,
					\label{eq:Btime}
				\end{align}
				and the $\GenL^+\left( \R^3 \right)$-right action
				\begin{equation}
						\left( \zeta, A
						\right)			\to \zeta \cdot 
							\begin{pmatrix}
								1 & 0 \\
								0 & A
							\end{pmatrix} 
						= \left( \zeta_0, A^i{}_1   \, \zeta_i ,
						 A^i{}_2   \, \zeta_i,
						 A^i{}_3   \, \zeta_i 
						\right) 
				\end{equation}
				for $\zeta = \left( \zeta_0,\zeta_1,\zeta_2,\zeta_3 \right)
				 \in \Frameb \left( T \mathcal Q \right)$ and
				$A \in \GenL^+\left( \R^3 \right)$. Then $\mathcal O$ 
				is a  $\GenL^+\left( \R^3 \right)$-reduction
				of the frame bundle $\Frameb \left(T \mathcal Q
				\right)$ with the property that there exists 
				a global frame field 
				$\xi \colon \mathcal Q 
				\to \Frameb \left( T \mathcal Q\right)$ 
				satisfying $\xi_0 = B$ and $\d \tau \cdot \xi_i =0$ for all
				$i \in \lbrace 1,2,3 \rbrace$, such that 
				\begin{equation}
					\mathcal O =\set{\zeta \in \Frameb \left(T \mathcal Q
				\right)}{\exists q \in \mathcal Q \, \exists 
							A \in \GenL^+\left( \R^3 \right) 
							\colon \zeta = \xi_q \cdot  
							\begin{pmatrix}
								1 & 0 \\
								0 & A
							\end{pmatrix} } \, .
				\end{equation}
		\item	\label{Item:Nconnec}
				The tangent bundle $T \mathcal Q$ is equipped with 
				a covariant derivative, called
				the \emph{Newtonian derivative $\nabla$}, which is 
	\begin{enumerate}[a)]
	\item compatible with the \emph{temporal metric $\d \tau^2 := 
	\d \tau \tp \d \tau$} : 	
		\begin{equation}
			\nabla \d \tau^2 = 0 \, , 
			\label{eq:nablatau}
		\end{equation}
	\item compatible with the spatial metric:	
		\begin{equation}
			\nabla \delta = 0 \, , 
			\label{eq:nabladelta}
		\end{equation}
	\item torsion-free, i.e. $\forall X,Y \in \mathfrak X 
	\left( \mathcal Q \right)$:  
		\begin{equation}
			\nabla _X Y = \nabla _Y X + \Lieb{X}{Y} \, .
		\end{equation}
	\end{enumerate} 
	\end{enumerate}
	The vector field $B$ is called the \emph{intrinsic 
	observer (vector) field}. 
	An (ordered) triple of tangent vectors $\left( Y_1, Y_2, Y_3 \right)
	$ at some $q \in \mathcal Q$ is called \emph{right-handed}, if  
	$( B_q, Y_1, Y_2, Y_3 ) \linebreak \in \mathcal O$. Analogously   
	we define right-handedness of a triple of vector fields. 
	Coordinates satisfying \eqref{eq:Eulercond} are called 
	\emph{Eulerian coordinates}, if in addition $( 
	\partial/\partial x^1,\partial/\partial x^2, \linebreak
	\partial/\partial x^3 )$ 
	is right-handed. 
	\end{subequations} 
\end{Definition}
For convenience, we identify the points 
$q \in \mathcal Q \subseteq \R^4$ with their Eulerian 
coordinate values, s.t. $q= \left( t, \vec x \right)$. 
Condition \ref{Item:metrics}) can be read as an integrability 
condition, i.e. the coordinates are chosen in accordance with 
the geometric structures and not vice versa. Thus the definition 
is coordinate-independent. The Newtonian orientation 
\ref{Item:Norient}) is 
necessary in the definition to be able to mathematically 
distinguish a physical system modeled on a Newtonian spacetime 
from its mirror image.  
It is easy to check that \eqref{eq:Bspace} and
\eqref{eq:Btime} uniquely determine $B$ to be 
\begin{equation}
	B = \partd{}{t} =: \partd{}{\tau} \, ,  
\end{equation}
so our definition of $\mathcal O$ is sensible. As it is the case 
for ordinary 
orientations on manifolds, there are precisely two possible 
Newtonian orientations $\mathcal O$ on $\mathcal Q$. 
\par
Clearly, the intrinsic observer field $B$ plays a special role. 
Condition \eqref{eq:Btime} means that the time form $\d \tau$ 
determines the parametrization of the integral curves of the 
intrinsic observer field, including its ``time orientation'', and 
condition \eqref{eq:Bspace} means that the integral curves of the 
observer field have no spatial length, or, equivalently, they 
describe mass points at rest. Therefore, due to the existence 
of a ``preferred rest frame'', \thref{Prin:Galileo} is actually 
violated in \thref{Def:Nst}, if one does not consider a 
Newtonian spacetime as the limiting case of a special relativistic 
model for a particular observer. Mathematically this is captured by 
the fact that Galilei boosts are not spatial isometries of a Newtonian 
spacetime, i.e. isometries with respect to the degenerate spatial 
metric. Within the special relativistic ontology, however, the 
Lorentz boosts are isometries of the physical spacetime and we can 
find a Newtonian spacetime corresponding to the boost by taking 
the Newtonian limit. This procedure yields 
two different spatial metrics, one for each observer. Therefore, 
\thref{Prin:Galileo} is indeed satisfied on an ontological level. 
\par
Excluding point \ref{Item:Nconnec}), Newtonian 
spacetimes trivially exist. The following lemma shows that the 
Newtonian connection is also well-defined.  
\begin{Lemma}[Existence \& Uniqueness of the Newtonian connection]
	\label{Lem:Nconn}
\begin{subequations}
	Let $\left( \mathcal Q , \d \tau, 
	\delta, \mathcal O \right)$ be a Newtonian spacetime. 
	Then the Newtonian 
	connection $\nabla$ is unique and trivial in Eulerian coordinates, 
	i.e. all the connection coefficients vanish.  
\end{subequations}
\end{Lemma}
\begin{Proof}
	Consider $g:= \d \tau^2 + \delta$. \eqref{eq:nablatau} 
	and \eqref{eq:nabladelta} in 
	\thref{Def:Nst} imply 
	\begin{equation}
		\nabla g = \nabla \left( \d \tau ^2 + \delta \right)
		= \nabla \d \tau^2 + \nabla \delta = 0 \, .
	\end{equation}
	Now $\nabla$ is just the Levi-Civita connection with respect to 
	the standard Riemannian metric $g$ in the global chart 
	$\left( \mathcal Q, x \right)$ and the result follows.  
\end{Proof}
\par
We conclude that our construction is both physically and 
mathematically consistent. Yet before we can set up physical models on a 
Newtonian spacetime $\left( \mathcal Q , 
\d \tau, \delta, \mathcal O \right)$, we need to consider the 
relevant dynamical quantities as obtained from the theory of relativity. 
These considerations will yield two subclasses of tangent vectors. 
\par
Recall that in relativity theory the spacetime model is a time-oriented%
\footnote{For the same reason as in the case of Newtonian spacetimes, 
it is sensible to assume spacetimes to be also space-oriented.} 
Lorentzian $4$-manifold $\left(\mathcal Q, g \right)$, 
equipped with the Levi-Civita 
connection $\nabla$. If a curve 
\begin{equation}
	\gamma \colon I \to \mathcal Q  
	\colon \tau \to \gamma \left( \tau \right) \, ,
\end{equation}
defined on an open interval $I \subseteq \R$ is assumed to describe 
physical motion, we require its tangent vector
field $\dot \gamma := \gamma_* (\partial/ \partial \tau)$ to be timelike, 
future directed and to be parametrized with respect to proper time $\tau$. 
For the latter 
	\begin{equation}
		g \left( \dot \gamma, \dot \gamma \right) = c^2
		\label{eq:cnorm} 
	\end{equation}
is a necessary and sufficient condition. Such curves $\gamma$ are known 
as observers and if for a tangent vector $X \in T \mathcal Q$ an observer 
$\gamma$ exists with $X= \dot \gamma_\tau$ for some $\tau \in I$, then $X$
is called an observer vector. Vector fields $X \in \mathfrak X 
\left(\mathcal Q \right)$ whose values $X_q \in T_q \mathcal Q$ are 
observer vectors at every $q \in \mathcal Q$ are accordingly called 
observer (vector) fields. Since a region of physical spacetime 
with negligible curvature can be approximately described by 
special relativity, we may restrict ourselves to the case where
$\mathcal Q \subseteq \R^4$ is open and $g= \eta$ is the Minkowski metric. 
In standard coordinates $\left( t, \vec x\right)$ on $\mathcal Q$, we  
write the tangent vector of an observer $\gamma$ as
	\begin{equation}
		\dot \gamma = \dot t \, \partial_t
		+ \dot x^i \, \partial_i \, ,
		\label{eq:Xonimage}
	\end{equation}
where the dot denotes differentiation with respect to $\tau$. 
On the other hand, condition \eqref{eq:cnorm} requires 
	\begin{equation}
		\dot t = \frac{1}{\sqrt{1- \left( 
		\frac{1}{c} \frac{\d \vec x}{\d t} \right)^2 }} \quad \, ,
		\label{eq:timedilat}
	\end{equation}
where we used the notation
\begin{equation}
	\left( \frac{\d \vec x}{\d t} \right)^2 := \delta \left( 
	\frac{\d}{\d t}, 
 \frac{\d}{\d t} \right) = \delta_{ij} \frac{\d x^i}{\d t} 
 \frac{\d x^j} {\d t}
 \, .
\end{equation}
A first order Taylor expansion of \eqref{eq:timedilat} in 
\begin{equation}
	\frac{1}{c}\abs{\frac{\d \vec x}{\d t}} := \frac{1}{c} \sqrt{
	\left( \frac{\d \vec x}{\d t} \right)^2 }
\end{equation}
around $0$ yields 
	\begin{equation}
		\dot t \approx 1 \quad , 
		\label{eq:approxt}
	\end{equation}
which is the expression for $\dot t$ in the Newtonian limit. This implies 
\begin{equation}
	\dot{\vec {x}} \equiv \frac{\d \vec x}{\d \tau} = \dot t 
	\frac{\d \vec x}{\d t} \approx \frac{\d \vec x}{\d t}  \quad .
	\label{eq:approxx}
\end{equation}
Plugging \eqref{eq:approxt} and \eqref{eq:approxx} back into 
\eqref{eq:Xonimage} we get 
	\begin{equation}
		\dot \gamma \approx  
		\partd{}{t} + \frac{\d  x^i}{\d t} \partd{}{x^i}  
			\quad .
		\label{eq:limdotgam}
	\end{equation}
If we carry this reasoning over to observer vectors 
$X \in T_q \mathcal Q$ at any $q \in \mathcal Q$, then we get in the 
Newtonian limit
	\begin{equation}
		X \approx \Evat{\partial_t}{q} + \vec X \, 
		\label{eq:X}
	\end{equation}
with $\vec X = X^i \Evat{\partial_i}{q}$. This is the reason 
for naming 
$B= \partial_t$ in \thref{Def:Nst} the `intrinsic observer vector 
field'. 
\par
To obtain the other important class of tangent vectors, we have 
a look at the dynamics. Hence we consider a test particle%
\footnote{Physically, a test particle is an almost point-like mass 
(relatively speaking), whose influence on the spacetime geometry can be 
neglected in the physical model of consideration.}%
, which is described by an observer $\gamma$ and has mass $m \in \R_+$.  
The force on the particle is defined by
	\begin{equation}
		F:= m \, 
		\frac{\nabla \dot \gamma}{\d \tau} \quad , 
		\label{eq:Newton}
	\end{equation}
which is just the generalization of Newton's second law to 
general relativity. 
Note that gravity is not a force, but a pseudo-force. 
Due to metricity of the connection and condition \eqref{eq:cnorm} we 
obtain 
	\begin{equation}
		g \left( \dot \gamma,  \frac{\nabla \dot \gamma}{\d \tau} 
		\right) = 0 \quad , 
	\end{equation}
which roughly means that the (relativistic) velocity is orthogonal to 
the (relativistic) acceleration. 
Applying this on \eqref{eq:Newton}, we get 
	\begin{equation}
		g \left( \dot \gamma , F \right) = 0 \quad ,
	\end{equation}
hence $F$ is spacelike \cite{O'Neill}*{Chap. 5, 26. Lemma}. In the 
Newtonian limit, the force field $F$ must stay ``spacelike''. This 
is indeed the case, which we see by using the definition 
\eqref{eq:Newton} of $F$ together with the approximation 
\eqref{eq:limdotgam} for $\dot \gamma$:
\begin{equation}
	\frac{F}{m} = \frac{\nabla \dot \gamma}{\d \tau} 
	\approx \frac{\nabla }{\d t} \left( 
	\partd{}{t} + \frac{\d  x^i}{\d t} \partd{}{x^i}   \right)  
	= \frac{\d^2 x^i}{\d t ^2 } \, \partd{}{x^i} \equiv 
	\frac{\d^2 \vec x}{ \d t ^2} \quad .
\end{equation}
This directly shows that $F^0$ has to vanish in the Newtonian limit. 
\par 
We have thus obtained the two types of tangent vectors 
(and hence curves and vector fields) 
of relevance in any physical model set in a Newtonian spacetime, 
i.e. tangent vectors $Y \in T \mathcal Q$ with either $Y^t=0$ or 
$Y^t=1$. Our discussion motivates the following definition. 
\begin{Definition}[Newtonian vectors]
	\label{Def:Nvf}
\begin{subequations}
	Let $\left( \mathcal Q , \d \tau, 
	\delta, \mathcal O \right)$ be a Newtonian spacetime. A tangent vector 
	$Y \in T \mathcal Q $ at $q \in \mathcal Q$ is called 
	\emph{Newtonian spacelike}, if $\d \tau \cdot Y= 0$ 
	or, equivalently, in Eulerian coordinates
	\begin{equation}
			Y =  \vec Y := Y^i \, \Evat{\partd{}{x^i}}{q} 
			 \, . 
	\end{equation}
	$Y \in T_q \mathcal Q $ is called a \emph{Newtonian observer vector}, 
	if $\d \tau \cdot Y= 1$ or, equivalently, 
	\begin{equation}
			Y = \Evat{\partd{}{t}}{q} +  Y^i \, 
			\Evat{\partd{}{x^i}}{q} = 
			\Evat{\partd{}{t}}{q} + 
			\vec Y \, .
	\end{equation}
	A tangent vector $Y$ is called \emph{Newtonian}, if $Y$ is 
	either a Newtonian observer vector or Newtonian spacelike. 
	For a Newtonian vector $Y$, we call $\vec Y$ the
	\emph{spacelike component of $Y$}. 
\end{subequations}
\end{Definition}  
It follows that a tangent vector 
$X$, describing the velocity vector of a point mass in the Newtonian limit 
at some instant, 
is a Newtonian observer vector, and a vector 
$F$, giving the force acting on such a particle 
according to \eqref{eq:Newton} 
at that instant, is Newtonian spacelike (i.e. $F = \vec F$). 
\begin{Remark}
	\label{Rem:Nvf}
\begin{subequations}
	The above terminology carries over to vector fields, e.g. a 
	\emph{Newtonian observer (vector) field $Y$} is one whose values 
	$Y_q \in T_q \mathcal Q$ are Newtonian observer vectors for every 
	$q \in \mathcal Q$. 
	We denote the space of (smooth) Newtonian vector fields by 
	$\mathfrak X_N \left( \mathcal Q \right)$, the space of (smooth)
	Newtonian spacelike vector fields by 
	$\mathfrak X_{Ns} \left( \mathcal Q \right)$ and the space
	of (smooth) Newtonian observer vector fields by 
	$\mathfrak X_{Nt} \left( \mathcal Q \right)$. 
	\par 
	Note that there are not any ``Newtonian lightlike'' 
	vectors. Indeed, for 
	physical consistency we require 
	$\lvert \vec X \rvert < c$. 
	\par 
	The space of Newtonian spacelike vector fields 
	forms a real vector space, the space of Newtonian observer 
	vector fields does not. However, if we add a Newtonian spacelike 
	vector field to a Newtonian observer vector field, we still have a 
	Newtonian observer vector field. The intrinsic observer field is then 
	the trivial Newtonian observer field, its integral 
	curves physically correspond to observers at rest with respect to 
	some inertial observer $\gamma$ in Minkowski spacetime $\left( 
	\R^4, \eta \right)$. 
\end{subequations}  
\end{Remark}
\par 
Instead of considering a single observer $\gamma$ in Minkowski 
spacetime, let us now assume that it is the integral curve of an 
observer field $X$. If each integral curve of $X$ describes the 
trajectory of a test particle of equal mass $m$, then \eqref{eq:Newton}
adapted to this case yields 
\begin{equation}
	F = m \nabla_X X \, .
	\label{eq:forcef}
\end{equation} 
In the Newtonian limit, we obtain the Newtonian spacelike 
vector field $F \approx \vec F$ and the Newtonian observer vector 
field $X \approx \partial_t + \vec X$, hence \eqref{eq:forcef} 
is approximated by 
\begin{equation}
	\vec F \approx m \nabla_{\partial_t + \vec X}
	\left( \partial_t + \vec X \right) = m \left( \partd{X^i}{t} + X^j 
	\, \partd{X^i}{x^j} \right) \, 	\partd{}{x^i} 
	\equiv m \biggl( \partd{\vec X}{t} + \nabla_{\vec X} \vec X
	\biggr)  
	\, .
	\label{eq:approxFfield}
\end{equation}
We thus see that the Newtonian limit naturally gives rise to what 
is known as the material derivative in the fluid mechanics 
literature \cite{Chorin}*{p. 4}. Intuitively, the material derivative 
of a Newtonian observer vector field $X$ along itself gives 
the acceleration of a point $\vec x$ in space moving along the 
flow lines of $X$ at some time $t$ \cite{Acheson}*{\S 1.2}. 
However, as we have obtained this from the Levi-Civita connection 
in the Newtonian limit and not in the context of fluids, we do not 
use this terminology here. Nonetheless we shall adapt our notation. 
So if $X \in \mathfrak X_{Nt} \left( \mathcal Q \right)$ is a 
Newtonian observer field and $Y \in \mathfrak X_N 
\left( \mathcal Q \right)$ a Newtonian vector field then, according 
to \thref{Lem:Nconn}, the Newtonian derivative $\nabla$ of $Y$ along 
$X$ can be written as
	\begin{equation}
		 \nabla _ X Y =  \partd{\vec Y}{t} 
		 + \nabla_{\vec X} \vec Y =:
		 \partd{\vec Y}{t} + \left( \vec X 
		 \cdot \nabla \right) \vec Y   \, 
	\end{equation}
in full compliance with \eqref{eq:approxFfield}. If 
$X \in \mathfrak X_{Ns} \left( \mathcal Q \right)$ is Newtonian spacelike 
instead, then 
	\begin{equation}
		 \nabla _ X Y = \nabla_{\vec X} \vec Y = 
		 \left( \vec X 
		 \cdot \nabla \right) \vec Y   \, .
	\end{equation}
This also shows that for Newtonian vector fields $X$, $Y$ the 
expression $\nabla _ X Y$ is always Newtonian spacelike. 
\par
For the special case of a Newtonian observer field $X$, 
we use the notation 
\begin{equation}
	\dot X := \nabla_X X =  \nabla_X 
	\vec X =  \dot{ \vec X } 
	= 
	\partd{\vec X}{t} 
		 + \left( \vec X \cdot \nabla \right) \vec X   
	\, ,
\end{equation}
which has the natural interpretation of acceleration. 
\par 
We still have to mathematically construct the relevant vector 
calculus operators on Newtonian spacetimes
without the need to refer to the Newtonian limit. 
\par
So let $\left( \mathcal Q , \d \tau, \delta, \mathcal O \right)$ 
be a Newtonian spacetime, define 
	\begin{equation}
		\Omega_t:= \set{\vec x \in \R^3}{\left( t, \vec x \right) \in 
		\mathcal Q} \quad , \quad 
		I := \set{ t \in \R}{\Omega_t \neq \emptyset}
		\label{eq:Omega}
	\end{equation}
and let 
$\iota_t \colon \Omega_t \to \mathcal Q$ be the natural inclusion. 
By the regular value theorem, 
there is a unique topology and smooth structure on $\Omega_t$ such that 
it becomes an embedded, smooth submanifold of $\mathcal Q$ and 
it can then be naturally equipped with the flat Riemannian 
metric $\iota^*_t \delta$. It also 
inherits a natural orientation from the Newtonian orientation 
$\mathcal O$ on $\mathcal Q$. Thus $\Omega_t$ is an oriented Riemannian 
$3$-manifold and hence the vector calculus operators $\grad$, $\div$ and 
$\curl$ are well defined (cf. \cite{Rudolph}*{Ex. 4.5.8} for definitions). 
These can be naturally extended to operators on 
$\mathcal Q$ by considering $T_{\vec x} \Omega_t$ as a linear subspace of 
$T_{\left( t, \vec x \right)} \mathcal Q$ 
for each $\left( t ,\vec x\right) \in \mathcal Q$. 
\begin{Definition}[Vector Calculus on Newtonian spacetimes]
	\label{Def:Nvc}
\begin{subequations}
	Let $\left( \mathcal Q , \d \tau, \delta, \mathcal O \right)$ be 
	a Newtonian spacetime, $X \in \mathfrak X _N 
	\left( \mathcal Q \right)$ 
	be a smooth Newtonian vector field, $f \in C^\infty \left( 
	\mathcal Q, \R \right)$ and let 
	$\iota_t \colon \Omega_t \to \mathcal Q$ be defined as above for each 
	$t\in \R$ such that $\Omega_t \neq \emptyset$. 
	We then define for every
	$\left( t, \vec x \right) \in \mathcal Q$ 
	\begin{enumerate}[i)]
		\item 	the \emph{gradient of $f$}, denoted by 
				$\nabla f \in \mathfrak X _{Ns} 
				\left( \mathcal Q \right)$, via 
				\begin{equation}
					\left( \nabla f \right) 
					_{\left( t,\vec x\right)} :=  
					\left(\grad \iota_t^*f\right)_{\vec x} \, ,
					\label{eq:Ngrad}
				\end{equation}
		\item 	the \emph{divergence of $X$}, denoted by 
				$\nabla \cdot X \in C^\infty
				\left( \mathcal Q, \R \right)$,  via
				\begin{equation}
					\left( \nabla \cdot X \right) \left( t, 
					\vec x \right) := \left( \div \left(
					{\vec{X}} _ {\iota_t \left( . \, \right)} 
					\right)  \right)
					\left( \vec x \right) \, ,
					\label{eq:Ndiv}
				\end{equation}
		\item	the \emph{curl of $X$}, denoted by 
				$\nabla \cross X  \in \mathfrak X _{Ns} 
				\left( \mathcal Q \right)$, via 
				\begin{equation}
					\left(\nabla \cross X \right)_
					{\left( t, \vec x \right)} := 
					\left( \curl 
					\left( 
					{\vec{X}} _ {\iota_t \left( . \, \right)} 
					\right) \right)_ 
					{ \vec x  }  \, , 
					\label{eq:Ncurl}
				\end{equation}
		\item	and the \emph{Laplacian of $f$} as
				\begin{equation}
					\Delta f := \nabla \cdot \left( \nabla f 
					\right).
					\label{eq:NDelta}
				\end{equation}
	\end{enumerate}  
\end{subequations}
\end{Definition}
Note that this definition just yields the ordinary 
vector calculus operators on $\R^3$, naturally adapted to the setting of 
Newtonian spacetimes. Similarly, the cross product $\cross$ can 
be extended from $T \Omega_t$ to $T\mathcal Q$. 
Moreover, the definitions naturally 
extend to complex valued functions and vector fields. 
\par
With this, we have finished our construction of a spacetime model, the 
associated (differential) operators and the elementary concepts needed for 
any physical model constructed upon it. 

\section{Local Equivalence of the Schr\"odinger and Madelung Equations}
\label{sec:equivalence}

We now employ the construction of the previous section to set up a 
model of a non-relativistic quantum system with one Schr\"odinger 
particle. 
\par 
In the Schr\"odinger picture of quantum mechanics 
\cite{Ballentine}*{\S 4.1 to \S 4.3} such a system under the 
influence of an external force
\begin{equation}
	\vec F = - \nabla V
\end{equation}
with potential $V \in C^\infty \left( \mathcal Q, \R \right)$ 
is described by a so called wave function 
$\Psi \in C^\infty \left( \mathcal Q, \C \right)$, satisfying the 
Schr\"odinger equation \cites{Schroedinger1,Schroedinger2,Schroedinger3}
\begin{equation}
	\i \hbar \partd{}{t}\Psi = - \frac{\hbar^2}{2m} \Delta \Psi + 
	V \Psi 
	\label{eq:SE}
\end{equation}
together with the rule that $\rho := \Psi^* \Psi \equiv 
{\abs{\Psi}}^2 $ gives the probability density for the particle's 
position at fixed time.%
\footnote{For a discussion on this interpretation and why 
alternative ones should be excluded, see e.g. \cite{Ballentine}*{\S 4.2}.}
This description has a number of disadvantages: 
\begin{enumerate}[i)]
	\label{enum:probs}
	\item 	The function $\Psi$ is complex and it is not apparent how and 
			why this is the case. This in turn prevents a direct 
			physical interpretation. 
	\item	The equation is already integrated, in the sense that it 
			is formulated in terms of the potential $V$ and 
			that the phase of $\Psi$ is only specified up to an arbitrary 
			real summand.
			This in turn suggests that the equation is not 
			fundamental, i.e. it is not formulated in terms of 
			directly measurable physical quantities. 
	\item	It is not apparent how to generalize the 
			Schr\"odinger equation to the case where no potential exists
			for a given force $\vec F$. 
	\item	It is not entirely apparent how to generalize the 
			Schr\"odinger equation 
			to more general geometries, i.e. what happens in the 
			presence of constraints, and what the underlying 
			topological assumptions are. 
	\item	Related to this is the fact that, 
			due to the $\partial \Psi / \partial t$ term, there is no 
			obvious relativistic generalization. This in 
			turn reintroduces the conceptual problems with 
			\thref{Prin:Galileo} on page \pageref{Prin:Galileo}.  
			\label{item:relgen}
	\item	Let $t \in  I$ and let
			$\mu_t$ be the canonical volume form on $\Omega_t$
			(cf. \eqref{eq:Omega}) 
			with respect to the 
			metric $\iota^*_t \delta$, i.e. $\mu_t = 
			\d x^1 \wedge \d x^2 \wedge \d x^3 \equiv \d^3x$. 
			The statement that for any Borel measurable 
			$N \subseteq \Omega_t \subseteq	\R^3$ the
			expression 
			\begin{equation}
				 \int_{N} \iota^*_t\rho \, \mu_t \quad \in 
				 \left[ 0,1\right]
			\end{equation}
			gives the probability for the particle to be found
			within the region $N$ at time $t$ is inherently 
			non-relativistic. Again this leads to problems 
			with \thref{Prin:Galileo}.
			\label{item:Born} 
\end{enumerate}
In this section we will observe that these problems are strongly 
related to each other and find their natural resolution in the 
Madelung picture.
\par 
Before we state and prove the main theorem of this section, that 
is \thref{Thm:equivalence}, we would like to remind the reader of 
the Weber identity 
\cite{Weber} known from fluid dynamics, 
since it is essential for passing between the Newtonian and 
the Hamiltonian description.
\begin{Lemma}[Weber Identity] 
	\label{Lem:Weber} 
	Let $\left( \mathcal Q , \d \tau, \delta, 
	\mathcal O \right)$ be a Newtonian spacetime and let $\vec X 
		\in \mathfrak X _{Ns} \left( \mathcal Q \right)$
	be a smooth Newtonian spacelike vector field.\\
	Then 
	\begin{equation}
		\left( \vec X \cdot \nabla \right) \vec X = \nabla 
		\left( \frac{\vec X^2}{2} \right) - 
		\vec X \cross \left( \nabla \cross \vec X \right) \, .
		\label{eq:Weber}
	\end{equation} 
\end{Lemma}
\begin{Proof} 
	Let $t \in I$ as defined in \eqref{eq:Omega}. For the 
	vector fields $\vec X^t := \vec X_{\iota_t \left( . \, \right)}, 
	\vec Y^t := \vec Y_{\iota_t \left( . \, \right)} \in 
	\mathfrak X \left(  \Omega_t \right)$ and the induced (standard)
	connection $\nabla$ on $\Omega_t \subseteq \R^3$, we have, using 
	standard notation, as a	standard result in vector 
	calculus in $\R^3$ (cf. \cite{Chorin}*{p. 165, Eq. 7}) that 
	\begin{equation*}  
		\nabla \left( \left(
		\iota^*_t \delta \right) \left(\vec X^t,\vec Y^t \right)\right)
		 = \vec X^t \cross \left( 
		\nabla \cross \vec Y^t \right) + \left(\vec X^t \cdot 
		\nabla \right) \vec Y^t +\vec Y^t \cross \left(\nabla \cross 
		\vec X^t \right) + \left( 
		\vec Y^t \cdot  \nabla  \right) \vec X^t \, .
	\end{equation*}
	To obtain \eqref{eq:Weber}, we set $\vec X^t = \vec Y^t$ and let 
	$t$ vary. 
\end{Proof}
\par 
We named \thref{Thm:equivalence} in the honor of Erwin Madelung, as 
it is mainly based on his article \cite{Madelung} and we merely 
formalized it to meet the standards of mathematical physics. Note that 
the choice of sign of $\varphi$ is pure convention. We choose it such 
that for $\partial/\partial t$ future directed in Minkowski 
spacetime $\left(\R^4, \eta \right)$ (cf. \cite{Kriele}*{Def. 3.1.3})
and $\partial \varphi / \partial t > 0$, the vector field 
\begin{equation}
	X = \frac{\hbar}{m} \grad \varphi \equiv \frac{\hbar}{m} 
	\, \eta^{-1} \cdot 
	\d \varphi 
\end{equation}
is future directed. 
\begin{Theorem}[Madelung's Theorem]
	\label{Thm:equivalence}
	\begin{subequations} 
	Let $\left( \mathcal Q , \d \tau, \delta, 
	\mathcal O \right)$ be a Newtonian spacetime, $m, \hbar \in \R_+$ 
	and let $I \subseteq \R$, $\Omega_t \subseteq{\R^3}$ be defined
	as in \eqref{eq:Omega}. \\
	If  $X \in \mathfrak X _{Nt} 
	\left( \mathcal Q \right)$ is a Newtonian observer vector field, 
	$\vec F \in \mathfrak X _{Ns} \left( \mathcal Q \right)$ 
	a Newtonian spacelike vector field, $\rho \in C^\infty 
	\left( \mathcal Q , \R_+ \right)$ a strictly positive, real 
	function and the first 
	Betti number $b_1 \left(\Omega _t \right)$ of $\Omega_t$
	vanishes for all $t \in I$, then 
	\begin{gather}
		m \dot{{X}} = \vec F + \frac{\hbar^2}{2m} \nabla 
		\frac{\Delta \sqrt{\rho}}{\sqrt{\rho}}  \, ,
		\label{eq:Madelung}\\
		\frac{\partial \rho}{\partial t}	+ \nabla \cdot 
		\left( \rho \, \vec X \right) = 0 \, , \label{eq:continuity} \\
		\nabla \cross \vec X = 0 \, , \label{eq:irrotX}\\
		\nabla \cross \vec F = 0 \, , \label{eq:irrotF}
	\end{gather}
	imply that there exist 
	$\varphi, V \in C^\infty\left( \mathcal Q, \R 
	\right)$ such that 
	\begin{gather}
		X = \partd{}{t} - \frac{\hbar}{m} \nabla \varphi \, , 
		\label{eq:gradvarphi} \\
		\vec F = - \nabla V \, , \label{eq:V} \\ 
		H :=  \frac{m}{2} \vec X ^2 + V - \hbar \partd {\varphi}{t} 
		-  \frac{\hbar^2}{2m} 
		\frac{\Delta \sqrt \rho}{\sqrt \rho} = 0 \, .  \label{eq:H}   
	\end{gather}
	Moreover, if one defines 
	\begin{gather}
		\Psi := \sqrt{\rho} \, e^{-\i \varphi} \, , \label{eq:Psi} 
	\end{gather}
	then it satisfies
	\begin{gather}
		\i \hbar \partd{}{t}\Psi = - \frac{\hbar^2}{2m} \Delta \Psi 
		+ V \Psi \, . \label{eq:Schroedinger}
	\end{gather}
	Conversely, if $\Psi 
	\in C^\infty\left( \mathcal Q, \C \setminus \lbrace 0 \rbrace 
	\right)$ and $V \in C^\infty \left( \mathcal Q, \R \right)$ satisfy 
	\eqref{eq:Schroedinger}, 
	define 
	$\rho := \abs{\Psi}^2 \in C^\infty \left( \mathcal Q , \R_+ \right)$, 
	$\vec F$ via \eqref{eq:V} and 
	\begin{equation}
	\vec X := \frac{\hbar}{m} \Im \left( 
	\frac{\nabla \Psi}{\Psi}\right)
	\equiv \frac{\hbar}{2 \i m} \left(\frac{\nabla \Psi}{\Psi}- 
	\frac{\nabla \Psi^*}{\Psi^*} \right)
	\label{eq:XfromPsi}
	\end{equation}
	such that $X:= \partial/ \partial t + \vec X$ is a Newtonian 
	observer vector 
	field. Then \eqref{eq:Madelung}, \eqref{eq:continuity}, 
	\eqref{eq:irrotX} and \eqref{eq:irrotF} hold.  
\end{subequations}
\end{Theorem}
\begin{Proof}
\label{Prf:equivalence}
\begin{subequations}
	``$\implies$'' 
	By the definition of curl \eqref{eq:Ncurl}, we have for any 
	fixed $t \in I$ 
	\begin{equation}
		\d \left( \left(\iota^*_t\delta\right) 
		\cdot \vec X_{\iota_t \left( . \, 
		\right)} \right) = 0 \quad , \, 
		\d \left( \left(\iota^*_t\delta\right) 
		\cdot \vec F_{\iota_t \left( . \, 
		\right)} \right) = 0 \, .
	\end{equation}
	Since $b_1 \left( \Omega_t \right) = 0$, all closed $1$-forms 
	are exact and  
 	hence $\exists \tilde \varphi^t, \tilde V^t
 	\in C^\infty \left( \Omega_t, \R \right)$:
 	\begin{equation}
		 \left(\iota^*_t\delta\right) \cdot \vec X_{\iota_t \left( . \, 
		\right)} = \d \tilde \varphi^t \quad , \, 
		\left(\iota^*_t\delta\right) \cdot \vec F_{\iota_t \left( . \, 
		\right)} = \d \tilde V^t \, .
		\label{eq:tildepots}
	\end{equation}
	If we now let $t$ vary and observe that 
	$\mathcal Q = \bigsqcup_{t \in I} 
	\Omega_t$, the left hand sides yield smooth $1$-forms on 
	$\mathcal Q$ and 
	so do the right hand sides. In other words, the function 
	\begin{equation}
		\tilde \varphi \colon \mathcal Q \to \R 
		\colon \left( t, \vec x \right) \to \tilde 
		\varphi^t \left( \vec x \right) =: 
		\tilde \varphi \left(t, x \right)
	\end{equation}
	has smooth partial derivatives $\partial \tilde \varphi / 
	\partial x^i$ 
	on $\mathcal Q$ for $i \in \lbrace 1,2,3 \rbrace$, but 
	$\partial \tilde \varphi / \partial t$ need not exist. However, if we
	integrate $\partial \tilde \varphi / \partial x^1$ with respect to 
	$x^1$, we obtain a smooth function on $\mathcal Q$, i.e. by 
	choosing the 
	integration constants appropriately we may assume 
	$\tilde \varphi \in C^\infty \left( \mathcal Q, \R\right)$. 
	We then repeat this argument to obtain $\tilde V 
	\in C^\infty \left( \mathcal Q, \R\right)$. 
	\par
 	Choosing $\varphi:= - m \tilde \varphi/ \hbar$ and $V := - \tilde V$, 
 	we get via \eqref{eq:tildepots} and 
 	\eqref{eq:Ngrad}, that \eqref{eq:gradvarphi} and
 	\eqref{eq:V} hold. 
 	\par 
 	Define now 
 	\begin{equation} 
		U := - \frac{\hbar^2}{2m} \,
		\frac{\Delta \sqrt{\rho}}{\sqrt{\rho}} \, ,
	\end{equation}
	and $\tilde U := V + U$. Using the Weber identity 
	(\thref{Lem:Weber}) together with \eqref{eq:irrotX}, equation 
	\eqref{eq:Madelung} reads
	\begin{equation}
		- \hbar \partd{}{t}\left( \nabla \varphi \right) + 
		\nabla\left( \frac{m}{2} \, \vec X^2 \right) = - \nabla 
		\tilde U \, .
	\end{equation}
	Due to smoothness of $\varphi$ and the Schwarz' theorem, we have 
	\begin{equation}
		\partd{}{t} \nabla \varphi = \nabla \partd{\varphi}{t}  \, ,
	\end{equation}
	and hence 
	\begin{equation}
		\nabla H \equiv \nabla \left( 
		\frac{m}{2} \, \vec X^2 + \tilde U  - 
		\hbar \partd{\varphi}{t} \right)= 0 \, .
	\end{equation}
	Thus $H$, as defined by the left side of \eqref{eq:H}, depends only 
	on $t$. If $H \neq 0$, we can redefine $V$ via 
	$V - H \to V$ as then $\vec F = - \nabla \left( V - H \right) = 
	- \nabla V$ remains true. Hence \eqref{eq:H} follows. 
	\par 
	We now define $\Psi$ via \eqref{eq:Psi}, $R:= \sqrt \rho$ 
	and calculate 
	in accordance with \eqref{eq:NDelta}:
	\begin{align}
		\Delta \Psi 
		&= \nabla \cdot \left( \nabla \left(
		R \, e^{-\i \varphi} \right) \right) \nonumber \\
		&= \nabla \cdot \left( \nabla R \, e^{-\i \varphi}
			- \i R \nabla \varphi \, e^{-\i \varphi} \right) \nonumber \\
		&= e^{-\i \varphi} \left( \Delta R - 2 \i \nabla R 
		\cdot \nabla \varphi - \i R \Delta \varphi - R 
		\left( \nabla \varphi \right) ^2 \right) \nonumber \\ 
		&= e^{-\i \varphi} \left( \Delta R - R \left( \nabla 
		\varphi \right)^2 - \i \left( 2 \nabla R \cdot \nabla 
		\varphi + R \Delta \varphi \right) \right)   \, .
		\label{eq:DeltaPsi}
	\end{align}
	Plugging $\rho = R^2$ and \eqref{eq:gradvarphi} into 
	\eqref{eq:continuity} yields 
	\begin{equation}
		2 R \partd{R}{t} - \frac{\hbar}{m} \left( 2 R  \, 
		\nabla R \cdot \nabla \varphi + R^2 \,  
		\Delta \varphi \right) = 0 \quad . 
		\label{eq:ImSE} 
	\end{equation}
	Since $R$ vanishes nowhere, 
	we can multiply with $m / (\hbar R)$, compare 
	with \eqref{eq:DeltaPsi} and arrive at 
	\begin{equation}
		- \Im \left( e^{ \i \varphi} \Delta \Psi \right) = 
		\frac{2m}{\hbar} \partd{R}{t} \, .
		\label{eq:ImDeltaPsi} 
	\end{equation}
	On the other hand, \eqref{eq:H} can also be reformulated in terms
	of $\varphi$ and $R$ to yield  
	\begin{equation}
		- \frac{\hbar^2}{2m} \left( \Delta R - R 
		\left( \nabla \varphi \right)^2 \right) - \hbar R 
		\partd{\varphi}{t} + V R = 0 \, .
		\label{eq:ReSE}
	\end{equation}
	By comparing this with \eqref{eq:DeltaPsi}, we see that we 
	can construct a $\Delta \Psi$ by adding 
	$\i$ times the imaginary part of $e^{\i \varphi} \Delta \Psi$ for 
	which we have the expression \eqref{eq:ImDeltaPsi}. This gives  
	\begin{equation}
		- \frac{\hbar^2}{2m} \Delta \Psi \, e^{ \i \varphi} +
		V R = - \frac{\hbar^2}{2m} \, \i \left(\frac{2m}{\hbar}
		\partd{R}{t} \right) + \hbar R \partd{\varphi}{t} 
		= \i \hbar \partd{R}{t} + \hbar R \partd{\varphi}{t} \, .
		\label{eq:almostSE}
	\end{equation}
	To take care of the right hand side, we notice 
	\begin{equation}
		\i \hbar e^{\i \varphi} \partd{\Psi}{t}	 = 
		\i \hbar \partd{R}{t} + \hbar R \partd{\varphi}{t} \, .
		\label{eq:partdtSE}
		\end{equation}
	Thus, by multiplying \eqref{eq:almostSE} by $e^{-\i \varphi}$, we 
	finally arrive at the Schr\"odinger equation 
	\eqref{eq:Schroedinger}. \\
	``$\impliedby $'' The reverse construction amounts to 
	Madelung's discovery \cite{Madelung}. We may define the real function
	$R := \abs \Psi =: \sqrt \rho$, yet, unfortunately, we cannot 
	write $\Psi$ as in \eqref{eq:Psi}, since the 
	complex exponential is not (globally) invertible. Instead we define
	$Q := \Psi / \abs \Psi$ and observe that by \eqref{eq:XfromPsi}
	\begin{equation}
		\vec X = \frac{\hbar}{m} \Im \left( 
			\frac{\nabla \left( R Q\right)}{ R Q} \right)
			= \frac{\hbar}{m} \Im \left( 
			\frac{\nabla  Q}{ Q} \right) \, .
			\label{eq:QtoX}  
	\end{equation} 
	We now do the calculation backwards with $Q$ instead of 
	$e^{- \i \varphi}$. So in analogy to \eqref{eq:DeltaPsi} we consider
	\begin{align}
		\Delta \Psi &= \nabla \cdot \left( 
		\nabla \left(R Q  \right) \right) 
		= \nabla \cdot \left( \nabla R \, Q + R \, \frac{\nabla Q}{Q} \, 
		Q\right) \nonumber \\
		&= \Delta R \, Q + 2 \, \nabla R \cdot \left( \frac{ 
		\nabla Q}{Q} \right) \, Q
		+ R \,  \nabla \cdot \left( \frac{ \nabla Q}{Q} \right) \, Q    
		+ R \, \left( \frac{ \nabla Q}{Q} \right)^2 \, Q \nonumber \\
		&= Q \left(\Delta R +  R \, \left( \frac{ \nabla Q}{Q} \right)^2 
		+ 2 \, \nabla R \cdot \left( \frac{ \nabla Q}{Q} \right) + 
		R \,  \nabla \cdot \left( \frac{ \nabla Q}{Q} \right) \right) 
		\label{eq:Q1}
	\end{align}
	and in analogy to \eqref{eq:partdtSE} we obtain 
	\begin{equation}
		\i \hbar \partd{\left( R Q\right)}{t} = 
		\i \hbar \partd{R}{t} Q + \i \hbar R \, 
		\left( \frac{\partd{Q}{t}}{Q}\right) \, Q \, \, . 
		\label{eq:Q2}
	\end{equation}
	Dividing the Schr\"odinger equation \eqref{eq:Schroedinger}
	by $Q$ and inserting \eqref{eq:Q1} as well as \eqref{eq:Q2}, we can 
	take the imaginary part $\Im$ as well as the real part $\Re$. This is 
	done by employing the facts that both commute with derivatives, 
	derivatives of $Q$ divided by $Q$ are purely imaginary and that 
	for any complex number $A \in \C$, we have 
	$\Re \left( \i A\right)= - \Im A$ and 
	$\Im \left( \i A\right) = \Re A$.  Then after some further 
	algebraic manipulation and using \eqref{eq:QtoX}, the imaginary 
	part yields 
	the continuity equation \eqref{eq:continuity} and the real part  
	gives \eqref{eq:H} with $\hbar\, \Im \left( (\partial Q / 
	\partial t) /Q \right)$ instead of 
	$- \hbar\, \partial \varphi / \partial t$. For the
	latter, we again use the Weber identity from \thref{Lem:Weber} and 
	note 
	\begin{equation}
		\nabla \left( \frac{\partd{Q}{t}}{Q} \right) 
		= \frac{\partd{}{t} \nabla Q}{Q} - \frac{\partd{Q}{t} 
		\nabla Q}{Q^2}
		= \partd{}{t} \left( \frac{\nabla Q}{Q}\right) \, .
	\end{equation}
	Recalling the definition \eqref{eq:V} of $\vec F$ we indeed 
	obtain \eqref{eq:Madelung}. Finally, \eqref{eq:irrotF} and 
	\eqref{eq:irrotX} are obtained by seeing that $\vec F$ is a gradient
	vector field and by calculating
	\begin{equation}
		\nabla \cross \left( \frac{\nabla Q}{Q}\right) 
		= \frac{\nabla \cross \nabla Q}{Q} - \frac{\nabla Q 
		\cross \nabla Q}{Q^2} = 0 \, .
	\end{equation}
	This completes the proof. 
\end{subequations}
\end{Proof}
Since for every $(t,\vec x ) \in \mathcal Q$ the open ball 
centered at the point is canonically a 
Newtonian spacetime as well, the theorem shows that
the Madelung equations for irrotational force fields 
and the Schr\"odinger equation are locally equivalent. 
\begin{Remark}[On the `Quantization Condition']
	\label{Rem:BSc}
\begin{subequations}
	In the literature one finds the claim that a 
	quantization condition needs to be added for the Schr\"odinger 
	and the Madelung 
	equations to be equivalent
	\citelist{\cite{Wallstrom0} \cite{Wallstrom} \cite{Takabayasi}*{\S 6} 
	\cite{Holland}*{\S 3.2.2}}, namely 
	\begin{equation}
		\frac{m}{2 \pi \hbar}	\oint_{\gamma} \iota_t^* \left( \delta
	\cdot X \right) \in \Z 
		\label{eq:BSc}
	\end{equation}
	for all $t \in I$ and all smooth loops $\gamma \colon 
	[0,2 \pi] \to \Omega_t$. 
	Note that, as observed by 
	Holland \cite{Holland}*{\S 3.2.2}, equation
	\eqref{eq:BSc} is astonishingly similar, yet inequivalent 
	to the Bohr-Sommerfeld quantization condition in the old quantum theory
	\cite{Sommerfeld}. Recalling Stoke's theorem 
	\cite{Rudolph}*{Thm. 4.2.14} and that 
	the irrotationality \eqref{eq:irrotX} of $X$ is equivalent to 
	closedness of $\iota_t^* \left( \delta
	\cdot X \right)$ for all $t \in I$, 
	we see that expression \eqref{eq:BSc} vanishes for $t \in I$ 
	and all $\gamma$
	if and only if
	$b_1 \left( \Omega_t\right) \equiv  0$. 
	\eqref{eq:BSc} can therefore only be
	relevant for the case $b_1 \left( \Omega_t\right) \neq 0$ for 
	some $t \in I$. Condition \eqref{eq:BSc} 
	originates from the simplest quantum mechanical model of 
	the Hydrogen atom and indeed excludes apparently unphysical 
	bound states, 
	but, as we will show in detail, is itself of topological origin. 
	\par
	We consider the Madelung equations for 
	a particle with charge $-q \in \left(- \infty, 0 \right)$ 
	being attracted 
	via the Coulomb force by a particle with charge $q$ fixed at 
	position $0 \in \R^3$.  
	The maximal domain where $\vec F$ is   
	smooth is $\R \cross \left(\R^3 \setminus 
	\left\lbrace 0 \right\rbrace \right)$, and 
	we have 
	$b_1 \left( \R^3 \setminus 
	\left\lbrace 0 \right\rbrace \right) = 0$. Together with 
	irrotationality \eqref{eq:irrotF} of $\vec F$, we can thus 
	find a potential 
	$V \colon \R \cross \left(\R^3 \setminus 
	\left\lbrace 0 \right\rbrace \right) \to \R$. 
	Moreover, 
	in spherical 
	coordinates  
	\begin{equation}
		\left(t,r, \theta, \phi \right) \colon \R^4 \setminus 
		\set{\left(t,\vec x\right) \in \R^4}{x^1 \geq 0, x^2=0}
		\to \R \cross \R_+ \cross \left(0, \pi \right) \cross 
		\left( 0, 2 \pi \right)
	\end{equation} 
	we can write the values of $V$ as $V \left( r\right)$, 
	since $\vec F$ is time-independent. 
	\par 
	If we now look for stationary (i.e. $t$-independent) 
	solutions of the Madelung equations, we find 
	the natural domains $\dom \rho = \dom X$ to be of the form
	$\R \cross \Omega =: \mathcal Q$ with open $\Omega \subseteq \R^3
	\setminus \lbrace 0 \rbrace$, but 
	in general we cannot assume $b_1 \left( \Omega \right) = 0$. 
	That is, to be able 
	to write down 
	the Schr\"odinger equation by application of 
	\thref{Thm:equivalence}, we have 
	to formally restrict ourselves to a (maximal, non-unique) 
	subset 
	$\Omega' \subseteq \Omega$ with $b_1 \left( \Omega' \right) = 0$. 
	The set $I \cross \Omega'$ is the natural domain of $\varphi$ 
	and $\Psi$, 
	but first we have to find the solution and then we 
	may fix 
	 $\Omega'$. 
	Due to the rotational symmetry of the problem, we 
	may already 
	assume 
	\begin{equation}
	\Omega' \subseteq W := 
	\R^3 \setminus \set{\vec x \in \R^3}{x^1 \geq 0,  x^2=0} 
	\subseteq \Omega 
	\end{equation} 
	such 
	that $\R \cross W = \dom \left( t,r, \theta, \phi \right)$. Note 
	that the assumption 
	of stationarity implies $\nabla (\partial \varphi/\partial t) = 0$, 
	but $\varphi$ may 
	be time dependent. If we now proceed, as usual, 
	by separation of variables in spherical coordinates, we obtain 
	a splitting 
	$\varphi \left( t, r ,\theta, \phi \right) = \varphi_0 \left( t\right)
	+\varphi_1 \left( r \right)+ \varphi_2 
	\left( \theta \right) + \varphi_3 
	\left( \phi \right)$ with $E \in \R$ 
	and 
	$\varphi_0 \left( t\right) = t E / \hbar$, 
	a radial equation and a spherical one. 
	The latter leads to 
	$\varphi_3 \left( \phi \right) = - \tilde m \phi$ 
	with $\tilde m \in \R$
	and the associated Legendre equation for 
	$\xi \colon (-1,1) \to \C \colon 
	y = \cos \theta \to \xi \left( y\right)$ given by
	\begin{equation}
		\left( 1- y^2 \right) \frac{\d^2 \xi}{ \d y^2} \left( y \right) 
		-2 y\frac{\d \xi}{ \d y} \left( y \right) + \left( l (l+1) 
		- \frac{\tilde m ^2}{1 - y^2}\right) \xi \left( y\right) = 0 \, . 
		\label{eq:Legendre}
	\end{equation}
	Now one usually asks for the condition 
	\begin{equation}
		\Psi \left(t, r, \theta, \phi \right) = 
		\Psi \left(t, r, \theta, \phi + 
		2 \pi k \right)
		\label{eq:globalpsi}
	\end{equation} 
	to be satisfied for some $x \in \mathcal Q$ and for all 
	$k \in \Z$, which constrains 
	$\tilde m$ (and ultimately the other quantum numbers $l$ and 
	$n$%
	) to be integer. 
	If $\Psi$ were a global function, \eqref{eq:globalpsi} 
	would follow from the continuity 
	of $\Psi$ on $\mathcal Q = \R \cross \Omega$ and the 
	property of $\Omega$, that there exists an 
	$\vec x \in \Omega$ such that the curve 
	$\gamma_{\vec x} \colon \R \to \Omega$, given by 
	\begin{equation}
		\gamma_{\vec x} \left( s \right):= 
		\left( \sqrt{\left( x^1\right)^2 + 
		\left(x^2 \right)^2} \cos s,\sqrt{\left( x^1\right)^2 + 
		\left(x^2 \right)^2} 
		\sin s , x^3 \right) \, ,
	\end{equation} 
	lies entirely in $\Omega$. 
	However, assumption \eqref{eq:globalpsi} cannot be made 
	if we only ask for $\Psi$ to be continuous on $\R \cross 
	\Omega' \subseteq 
	\R \cross W$. As equation 
	\eqref{eq:Legendre} also admits solutions for $l, \tilde m$ 
	not integer 
	\citelist{\cite{MacRobert}*{p. 288ff} \cite{Hobson}*{p. 180f}}, we 
	may continue to solve the other equation%
	\footnote{Please note that only the $(E<0)$-solutions are 
	admissible, as the other ones are not $L^2$-integrable 
	(c.f. \cite{Landau1}*{\S 36}).}
	and ultimately 
	find that there are solutions $\Psi$ with 
	$\tilde m \notin \Z$ and $X 
	\in \mathfrak X \left( \R \cross \Omega' \right)$, 
	given by 
	\begin{equation}
		X_{\left(t, r, \theta, \phi \right)} = 
		\Evat{\partd{}{t}}{\left(t, r, \theta, \phi \right)} +
		\frac{\hbar \, \tilde m}{m r^2  
		\sin^2 \theta} \, \Evat{\partd{}{\phi}}{\left(t, r, 
		\theta, \phi \right)} \,
		\label{eq:hydroX}
	\end{equation}
	in spherical coordinates. Please keep in mind that $\Omega'$ 
	also depends 
	on $\abs \Psi$, in particular we have to exclude all zeros 
	of the wave function. 
	Yet the field $X$, as given 
	by \eqref{eq:hydroX}, can be smoothly extended to 
	$\R \cross W'$ with 
	\begin{equation}
		\Omega' \subset W' := \R^3 \setminus 
		\set{\vec x \in \R^3}{x^1 = x^2=0} \subseteq \Omega
	\end{equation} 
	and $b_1 \left( W' \right) = 1$. For this $X$ equation 
	\eqref{eq:BSc} does not hold. \eqref{eq:BSc} 
	would indeed hold for all stationary solutions, 
	had we ad hoc assumed that 
	$\Psi$ is a global function, i.e. $\Omega' = \Omega$. Conversely, 
	had we ad hoc assumed 
	condition \eqref{eq:BSc}, then $\Psi$ would be a 
	global function. 
	\par 
	We conclude that the Madelung equations in general admit 
	more solutions than the Schr\"odinger equation, if the latter 
	is assumed to be globally valid. 
	However, since any point in $\mathcal Q \subseteq \R^4$ admits a 
	contractible neighborhood, \thref{Thm:equivalence} shows 
	that the claim that ``theories 
	based on the Madelung equations simply do not reproduce 
	the Schr\"odinger equation'' \cite{Wallstrom}*{\S IV} 
	is incorrect. While the 
	existence of apparently unphysical additional solutions 
	in this model of the 
	hydrogen atom does indicate a potential defect of the model, it 
	does not imply that the 
	Madelung equations yield an incorrect description of 
	quantum phenomena: This model 
	of the hydrogen atom neglects the motion of the nucleus, 
	the dynamics of   
	the electromagnetic fields, as well as relativistic effects. 
	It is thus plausible that 
	the problem expressed in \citelist{\cite{Wallstrom0} \cite{Wallstrom} 
	\cite{Takabayasi}*{\S 6} \cite{Holland}*{\S 3.2.2}}
	stems from an oversimplification of the physical situation. 
	Moreover, Wallstrom raised the interesting question of stability 
	of stationary solutions in this model \cite{Wallstrom}*{\S IV}. 
	Since unstable solutions are in a sense `unphysical', 
	it might be possible to exclude the additional ones on that 
	ground. 
\end{subequations}  
\end{Remark}
\par 
We now fix some terminology, that is partially derived 
from \cite{Wallstrom} and partially our own. 
The \emph{Madelung picture} consists of the 
\emph{Madelung equations}, that is
\begin{enumerate}[i)]
	\item the \emph{Newton-Madelung equation} \eqref{eq:Madelung}, 
	\item the continuity equation 
	\eqref{eq:continuity}, 
	\item 	the vanishing vorticity/irrotationality of the 
			\emph{drift (velocity) field} 
			$X$ \eqref{eq:irrotX}, 
\end{enumerate}
the topological condition $b_1 \left( \Omega_t \right) = 0$ for all 
$t \in \R$ and the irrotationality of the (external) force 
\eqref{eq:irrotF}. 
Obviously, the Madelung equations are a system 
of partial differential equations of third order in 
the \emph{probability density} $\rho$ and of first order in the 
drift field $X$. 
That means in particular, that $\rho$ and $X$ are the primary 
quantities of interest in the Madelung picture, as opposed to e.g. 
(time-dependent) wave functions in the Schr\"odinger picture or 
(time-dependent) operators in the Heisenberg picture. It is 
therefore justified to call a solution of the Madelung equations 
$\left( \rho, X \right)$ a \emph{state (of the system)} 
and $\left( \rho_t, \vec{X}_{t} \right)$ with $\rho_t \in C^\infty 
\left(\Omega_t, \R_+\right)$, $\vec{X}_{t} 
\in \mathfrak X \left( \Omega_t \right)$
a \emph{state (of the system) 
at time $t \in I$}. 
The flow of the drift field is called the \emph{drift flow} 
or \emph{probability flow} and the mass of the particle times 
the drift field is called the \emph{drift momentum field}, for 
reasons explained in section \ref{sec:interpretation} on page 
\pageref{sec:interpretation}. 
The drift field $X$ is a Newtonian observer vector field and, 
in accordance with \thref{Def:Nvf}, $\vec X$ is the spacelike 
component of the drift field.   
In reminiscence of the hydrodynamic analogue (unsteady potential flow) 
\cite{Chorin}*{\S 2.1}, we call \eqref{eq:H} the 
\emph{Bernoulli-Madelung equation}. The 
operator $U \colon C^\infty \left( \mathcal Q, \R_+\right) \to C^\infty 
\left(\mathcal Q, \R \right)$, as defined by
\begin{equation}
	U \left( \rho \right) := - \frac{\hbar^2}{2m} 
		\frac{\Delta \sqrt{\rho}}{\sqrt{\rho}} \quad , 
\end{equation}
is known as the \emph{quantum potential} or 
\emph{Bohm potential}. Analogously, we call 
the operator $\vec F _B := -\nabla U \colon 
C^\infty \left(\mathcal Q, \R_+ \right) 
\to \mathfrak X _{Ns} \left( \mathcal Q\right)$ 
(with $\left(\nabla U\right) \left( \rho\right) 
:= \nabla \left( U \left( \rho \right)\right)$)
the \emph{quantum force} or \emph{Bohm force}. This terminology 
is primarily historically motivated, we 
emphasize that the 
interpretation of $-\nabla U$ as an 
actual force is deeply problematic. Again, we refer to 
section \ref{sec:interpretation} on page 
\pageref{sec:interpretation}. 
\par
A priori, there are four real-valued functions constituting 
a solution of the Newton-Madelung equation: $\rho$ and 
three components of $\vec X$. If $X$ is 
irrotational and $b_1 \left( \Omega_t\right) \equiv 0$, it is 
enough to know 
the two functions $\rho$ and $\varphi$ (or 
the wave function $\Psi$) 
to fully determine 
the physical model. If a solution $\Psi$ of the Schr\"odinger equation is 
known, the simplest way to recover $\rho$ and $X$ is by calculating
\begin{equation}
	\rho = \Psi^* \Psi
\end{equation} 
and the spacelike component $\vec X$ of $X$ via 
\eqref{eq:XfromPsi} on page 
\pageref{eq:XfromPsi}. 
So by using Madelung's theorem (\thref{Thm:equivalence}), we 
can move freely between
the Schr\"odinger and Madelung picture, at least locally. 
\begin{Remark}[On time dependence]
	\label{Rem:time}
\begin{subequations}
	In correspondence with the arguments outlined in 
	section \ref{sec:Newton} on page \pageref{sec:Newton}, the 
	irrotationality of $X$ is a consequence of 
	the (special-)relativistic condition 
	\begin{equation}
		\d \left( \eta \cdot X \right) = 0 \, , 
	\end{equation}
	where $X$ is an observer vector field on an open 
	subset of Minkowski spacetime $\left( \mathcal Q
	\subseteq \R^4, \eta \right)$. As noted before, in the Newtonian 
	limit $X^0 \approx c$ and we thus obtain the conditions 
	\begin{equation}
		\frac{1}{c}
		\partd{\vec X}{t} \approx 0 , \quad \nabla \cross \vec X = 0
	\end{equation} 
	instead of mere irrotationality on $\mathcal Q$ to stay 
	consistent within the relativistic ontology. 
	That is, if $X$ is irrotational, it must also be approximately 
	time-independent in the above sense or the (naive) 
	Newtonian limit breaks down. 
\end{subequations}
\end{Remark}
\par 
There is a mathematical problem that deserves to be mentioned.
\begin{Question}[Existence and Uniqueness of Solutions]
	\label{Que:existence}
	Assuming that the probability density $\rho$ and the drift 
	field $X$ are given 
	and smooth on $\Omega \equiv \Omega_0$, under which 
	conditions does there 
	exist a smooth solution to the Madelung equations? 
	Is it unique? Is the vector field $X$ complete? 
\end{Question}
Apparently the question has been partially resolved by J\"ungel et al. 
\cite{Juengel}, 
who showed local existence and uniqueness of weak solutions in the 
special case of $X$ being a gradient vector field.  
\par
Returning to our original discussion in the beginning of this section, 
how do the Madelung equations offer a resolution of the 
problems associated with the Schr\"odinger equation 
addressed on page \pageref{enum:probs}? 
\par
We see that the use of the complex function $\Psi$ makes it 
possible to rewrite the Newton-Madelung equation and the 
continuity equation into one complex, second order, linear 
partial differential equation, which is arguably simpler to solve. 
Thus one can view the Schr\"odinger equation as an intermediate 
step in solving the Madelung equations, as has already been noted 
by Zak \cite{Zak}.  
The Madelung equations are formulated in terms of 
quantities that do not have a ``gauge freedom'', that means all 
the quantities in the Madelung equation are in principle uniquely 
defined and physically measurable. This argument alone 
is sufficient to consider the Madelung equations more fundamental 
than the Schr\"odinger equation. 
For example, the actual physical quantity corresponding to the 
phase $\varphi$ must be a coordinate-independent derivative 
thereof, as the physically measurable predictions in the 
Schr\"odinger picture are invariant under the transformation 
$\varphi \to \varphi + \varphi_0$ with $\varphi_0 \in \R$ and, 
of course, coordinate transformations.  A similar argument can 
be made for the potential $V$ of the force $\vec F$. 
Thus, if one wishes to generalize the description of quantum systems 
with one Schr\"odinger particle to the relativistic and/or 
constrained case, starting with the Madelung equations rather than the 
Schr\"odinger equation is the natural choice. Indeed, the 
Madelung equations offer a straight-forward (though unphysical, 
cf. footnote \ref{ftn:KG}) 
generalization of 
the Schr\"odinger equation to the (general)-relativistic case, 
but we will not discuss this here. For this reason, the resolution 
of \ref{item:relgen}) and \ref{item:Born}) will be postponed. 
The treatment of constrained non-relativistic systems can be 
approached by either 
solving the Madelung equations together with these constraints directly 
(cf. \S\ref{sec:interpretation1} for the interpretation of 
$\rho$ and $X$) or by
passing over to a Hamiltonian formalism with the use of the 
Bernoulli-Madelung equation \eqref{eq:H}. A generalization 
of the Madelung equations for non-conservative forces is 
immediate and the generalization to dissipative 
systems has been pursued in \cite{Tsekov1}. Note that 
for some generalizations 
it might not be possible to construct a Schr\"odinger equation, notably 
for rotational drift fields and forces.
\begin{Remark}[Geometric constraints]
	\label{Rem:gconstr}
	If the constraint is geometric, i.e. if the 
	particles are constrained to an embedded submanifold $M$ of 
	$\R^3$, like the surface of a sphere or a finite M\"obius band, 
	the adaption 
	of the Madelung equations follows the same precedure as for any 
	other Newtonian continuum theory: 
	\begin{enumerate}[i)]
	\item 	We first assume that $\mathcal Q$ is an open subset of 
			$\R \cross M \subseteq \R^4$ instead of 
			$\R^4=\R \cross \R^3$ and 
			pull back the structures 
			$\d \tau, \delta, \nabla$ on $\R^4$ via 
			the inclusion map 
			$\xi \colon \R \cross M \to \R \cross \R^3$. 
			Defining $I, \Omega_t, \iota_t$ as in \eqref{eq:Omega} 
			and taking again the pullback of the spatial metric 
			to get $h_t$
			for each $t \in I$, 
			this yields the vector calculus operators 
			divergence, gradient 
			and Laplacian on $\mathcal Q$, in full analogy to 
			\thref{Def:Nvc}. 
	\item 	Then we write down the continuity equation for	
			these new vector calculus operators, $\rho \in C^\infty
			\left(\mathcal Q ,\R \right)$ and $X= \partial/ \partial t 
			+ \vec X \in \mathfrak X 
			\left(\mathcal Q \right)$ such that 
			$\vec X_{\iota_t}$ is tangent to $\Omega_t$ for every 
			$t \in I$ (defined in full analogy to 
			\eqref{eq:Omega}). 
	\item 	Restrict the force to $\mathcal Q \subseteq \R \cross M$ 
			and take 
			only the tangential parts, then 
			write down the Newton-Madelung equations for the new force 
			$\vec F$ and vector calculus operators. 
	\item 	We replace the irrotationality of $\vec X$ by the condition 
			\begin{equation}
				\d \left( h_t \cdot \vec X_{\iota_t}\right) = 0 
				\quad \quad \forall t \in I \, .
			\end{equation}
	\item	To construct a Schr\"odinger equation, we require that the 
			new $\vec F$ also satisfies the above condition 
			and, of course, 
			the topological condition $b_1 \left(\Omega_t  \right)= 0$ 
			for all $t \in I$ needs to hold. Then proceed as in the proof 
			of \thref{Thm:equivalence}. 
	\end{enumerate}
	We conjecture that 
	this procedure just yields the ordinary Schr\"odinger equation 
	on $\mathcal M$ with Laplacian induced by $h$ (considered as 
	Riemannian for fixed $t$). 
\end{Remark}   
\par 
Madelung's theorem also gives an explicit condition for the 
global equivalence of the equations, which is, of course, 
topological. 
However, in practice we do not know the natural domain 
$\mathcal Q$ of $\rho$ and $X$ in advance, but we are given 
(sufficiently smooth) 
initial values of $\rho$ and $X$ on $\lbrace 0 \rbrace \cross \Omega$ 
with $\Omega = \Omega_0 \subseteq \R^3$ and would
then like to know whether we can apply \thref{Thm:equivalence} globally. 
There is a convenient answer to this question by noting that, on 
the grounds of 
\thref{Thm:interpretX} on page \pageref{Thm:interpretX}, we may 
identify $\mathcal Q \subseteq \R^4$ to be the image of 
$\lbrace 0 \rbrace \cross \Omega$ 
under the flow of $X$. Since we would like to have a global 
dynamical evolution of the system, we may assume that there 
exists an open interval $I \subseteq \R$ such that the 
flow $\Phi_t$ of $X$ is defined for all $t \in I$. The next proposition 
states the topological consequences of this situation. 
\begin{Proposition}[Topology of $\Omega_t$ and $\mathcal Q$]
	\label{Prop:topOQ}
\begin{subequations}
	Let $\left( \mathcal Q , \d \tau, \delta, \mathcal O \right)$ 
	be a Newtonian 
	spacetime, $I \subseteq \R$ be an open interval with $0 \in I$
	and let $\Omega := \Omega_0 = 
	\set{\vec x \in \R^3}{\left( 0, \vec x\right)
	\in \mathcal Q}$. Further, let $X$ be a Newtonian observer 
	vector field with flow 
	$\Phi$, such that $\mathcal Q$ is the image $\Phi_{I}\left( 
	\lbrace 0 \rbrace \cross \Omega\right)$.  \\ 
	Then $\Omega_t := \set{ \vec x \in \R^3 }{ (t, \vec x) \in \mathcal Q}$
	is diffeomorphic to  
	$\Omega$ and $\mathcal Q$ is diffeomorphic to $I \cross \Omega$. 
	In particular, the Betti numbers $b_i \left( \Omega\right)$, 
	$b_i \left( \Omega_t\right)$ and 
	$b_i \left( \mathcal Q \right)$ coincide for every 
	$i \in \N_0 := \N \cup \lbrace 0 \rbrace $ and $t \in I$. 
\end{subequations}
\end{Proposition}
\begin{Proof}
\begin{subequations}
	Define $\phi := \Phi \evat{I \cross 
	(\lbrace 0 \rbrace \cross \Omega)}$. 
	Hence $\dom \phi = I \cross \Omega$ and $\phi$ is surjective onto 
	$\mathcal Q$. Since $X$ is a Newtonian observer vector field, we find 
	that there exists a smooth function $\vec{\Phi} \colon 
	I \cross \Omega \to \mathcal Q$ such that for all 
	$t \in I, \vec x \in \Omega$: 
	\begin{equation}
		\phi \left( t, \vec x\right) = \Phi_t \left(0, \vec x \right)= 
		\bigl(t,  \vec{\Phi}_t \left( \vec x \right) \bigr) .  
	\end{equation}
	Since $\Phi_t$ is injective for any $t \in I$, so is 
	$\vec \Phi_t \colon \Omega \to \Omega_t$ and also $\phi$. As
	$\phi \left(t, . \, \right) = \Phi_t \left( 0 , . \, \right)
	= \left( t, \vec \Phi_t  
	\left( .\, \right)\right)$ 
	for all $t \in I$, the differential 
	$(\partial \vec \Phi_t / \partial \vec x)$ 
	of $\vec \Phi_t$ has full rank 
	(cf. \cite{Rudolph}*{Prop. 3.2.10/1}) and thus 
	\begin{equation}
		\phi_* = 
		\begin{pmatrix}
			1	&	0		 \\ 
			\partd{\vec \Phi}{t} & \partd{\vec \Phi}{\vec x}
		\end{pmatrix}
	\end{equation}
	has full rank. Since a smooth bijection whose differential 
	has full rank everywhere 
	is a diffeomorphism, $\phi$ is a diffeomorphism. Therefore $\mathcal Q$
	is diffeomorphic to $I \cross \Omega$ and since $\Omega_t$ 
	is an embedded 
	submanifold of $\mathcal Q$, it is diffeomorphic to 
	$\lbrace t \rbrace \cross 
	\Omega$ under $\phi$ and hence diffeomorphic to $\Omega$ itself. 
	\par
	Since diffeomorphic manifolds are (smoothly) homotopy equivalent, 
	$\mathcal Q$ is homotopy equivalent to $I \cross \Omega$ 
	and all $\Omega_t$s
	are homotopy equivalent to $\Omega$.
	One can directly proof from the definition of (smooth) 
	homotopy equivalence 
	(see e.g. \cite{Rudolph}*{Def. 4.3.5}) that for any smooth manifold 
	$\Omega$ and an open interval $I$, the product $I \cross \Omega$ is 
	homotopy equivalent to $\Omega$. Thus $\mathcal Q$ is also 
	homotopy equivalent 
	to $\Omega$. Since homotopy equivalent manifolds have 
	isomorphic de Rham 
	cohomology groups (cf. \cite{Rudolph}*{Cor. 4.3.10}), their 
	Betti numbers coincide.   
\end{subequations}
\end{Proof}
This means that under physically reasonable assumptions, the 
global applicability of \thref{Thm:equivalence} is determined by 
the topology of the initial value hypersurface $\Omega$. 
If one works in the relativistic ontology, this condition 
$b_1 \left( \Omega_t \right) = 0$ for all $t \in I$ 
should be replaced by $b_1 \left( \mathcal Q \right) = 0$. 
In the (naive) Newtonian limit, \thref{Prop:topOQ} then states 
that the latter condition implies the former one. We also wish 
to note that, if $\Omega$ is not connected, \eqref{Prop:topOQ} 
prevents the components from `merging' - in the sense 
that a solution cannot be extended to later times. 
Physically, this means that a two-particle model is more 
appropriate in this situation. 
\par 
If the condition $b_1 \left( \Omega_t\right) 
\equiv 0$ is not satisfied (as in \thref{Rem:BSc}), 
a global $\varphi$ need not exist and as a consequence 
a global wave function cannot be constructed. 
Apart from stationary solutions on $\mathcal Q = \R \cross \Omega$ with 
$b_1 \left( \Omega \right)\neq 0$, 
this can happen, for example, when attempting to 
describe the Aharonov-Bohm effect \cite{Aharonov}, or when a 
connected component of the  domain of the initial probability 
density $\rho_0 := \rho \left(0, \, \right)$ is not simply connected. 
It is unknown to us whether topological problems, as expressed in 
\thref{Rem:BSc} on page \pageref{Rem:BSc}, also occur in other 
quantum mechanical models. 
If not, it is possible to argue that the Madelung equations are the 
global version of the Schr\"odinger equation. This might 
yield additional physical solutions. 

\section{Relation to the Linear Operator Formalism} 
\label{sec:operator}

The Schr\"odinger picture of quantum mechanics is not  
limited to the Schr\"odinger equation, but also gives a set of 
rules how to determine expectation values, standard deviations 
and other probabilistic quantities for physical observables like 
position, momentum, energy, et cetera. In this section we 
examine how the Schr\"odinger picture and the Madelung picture 
relate to each other. We will observe that the Madelung picture 
suggests modifications of the current axiomatic framework of 
quantum mechanics, namely the replacement of the von Neumann 
axioms with the axioms of standard probabilty theory by 
Kolmogorov. 
Again, we will restrict ourselves to the 
$1$-particle Schr\"odinger theory, but our 
treatment has consequences for 
the general axiomatic framework of quantum mechanics. 
\par 
The general, mathematically naive formalism of quantum mechanics 
states \citelist{\cite{Hall}*{\S 3.6} \cite{Ballentine}*{\S 2.1}}, 
that for every ``classical observable'' $A$ there is a 
linear mapping%
\footnote{Note that this can already not be the case for the 
momentum operator $\hat{\vec {p}}$, but can only hold true for its 
``components'' $\hat p_i$. 
} 
 $\hat A \colon \mathcal H \to 
\mathcal H$ of some Hilbert space $\mathcal H$ with 
inner product $\inp{.}{.}$, that is assumed to be 
hermitian/self-adjoint, 
in the sense that $\forall \Psi, \Phi \in \mathcal H$ the 
operator $\hat A$ satisfies: 
\begin{equation}
	\inp{\Psi}{\hat A \Phi} = \inp{\hat A\Psi}
	{\Phi} \, .
	\label{eq:hermiticity}
\end{equation} 
The self-adjointness \eqref{eq:hermiticity} assures that 
the eigenvalues of $\hat A$, if they exist, are real.  This is 
necessary, because the eigenvalues are taken to be the values of 
the observable $A$ and these have to be real, physical quantities. 
Note that the time $t$ is treated as a parameter in this formalism 
and both $\hat A$ and $\Psi$ may depend on it. 
\par
In the Schr\"odinger theory, $\mathcal H$ is assumed to be 
a vector space consisting of functions $\Psi$ from some open subset 
$\Omega$ of $\R^3$ to $\C$. Moreover, it should be equipped 
with the $L^2$-inner product \cite{Becnel}*{\S B}
\begin{equation}
	\inp{.}{.} \colon \quad \mathcal H \cross \mathcal H 
	\to \C \quad \colon \quad  \left(\Psi, \Phi \right) \to 
	\inp{\Psi}{\Phi} := \int_{\Omega} \d^3 x \, 
	\Psi^*\left( \vec x \right) \Phi\left( \vec x \right) \,  \, . 
\end{equation}
Hence $\mathcal H$ ought to be a linear subspace of 
$L^2 \left( \Omega, \C \right)$ and, to assure completeness, it 
ought to be closed. The fact, that this formalism 
does not allow for the domain of $\Psi$ to change over time, can 
be remedied by allowing $\Omega$ and hence $\mathcal H$ to be 
time-dependent, but then one does not just have one 
Hilbert space, but a collection 
\begin{equation}
	\set{\mathcal H_t\, }{\, t \in I \subseteq \R} 
	\quad \text{with} \quad \mathcal H_t 
	\subseteq L^2 \left( \Omega_t, \C \right) \quad 
	\text{for all} \quad t \in I \, .
\end{equation}
So the formalism of Newtonian spacetimes is also implicitly 
used in this approach. 
\par  
The most common operators in the Schr\"odinger theory are the 
position operators $\hat x^i = x^i$, the momentum operators 
$\hat p_i = \i\hbar \, \, \partial/\partial x^i$, 
the energy operator%
\footnote{Usually this is called the Hamiltonian operator, but 
we take the Hamiltonian to be \eqref{eq:H}. Considering 
$\i \hbar \, \partial/\partial t$ as the energy operator is more 
natural from the relativistic point of view. } 
$\hat E = \i \hbar \partial_t$ and 
the angular momentum operators $\hat L_i = \varepsilon_{ij}{}^k \hat{x}^j 
\hat p_k$. With the exception of the position operator, these are all
related to spacetime symmetries (cf. \cite{Ballentine}*{\S 3.3}), and 
they are the operators that are used to heuristically construct 
more general ones by consideration of the ``classical analogue''. 
The question, which other observables are admissible, is in general not 
answered by the formalism itself - a problem which was used to justify 
more sophisticated quantization algorithms \citelist{
\cite{Woodhouse}*{\S 8.1} \cite{Waldmann}*{\S 5.1.2}}. 
We leave the Galilei operators and spin operators aside in this article, 
as the former are better treated in the context of approximate 
Lorentz boosts and spin is not covered here. 
\par 
The knowledge of the operators ${\hat {x}}^i$ and 
${\hat{p}}_j$ is enough to show the naivet\'e of the 
Hilbert space formalism: Assuming a sufficient degree of 
differentiability of functions in $\mathcal H$, 
we obtain the canonical commutation relation 
\begin{equation}
	\Lieb{\hat {x}^i}{{\hat{p}}_j} = \i \hbar \, \delta^i_j \, .
\end{equation}
It is common knowledge among mathematical physicists that 
this is in direct conflict with the self-adjointness of 
$\hat {x}^i$ and $\hat {p}_j$. For the sake of coherence, 
we state and prove the relevant assertion.  
\begin{Proposition}
	\label{Prop:Neumann}
	There does not exist any Hilbert space
	$\left( \mathcal H, \inp{.}{.} \right)$ with 
	linear maps $\hat x, \hat p 
	\colon \mathcal H \to \mathcal H$ such that
	the following hold: 
	\begin{enumerate}[i)]
		\item 	$\hat x, \hat p$ are self-adjoint. 
		\item $\hat x, \hat p $ satisfy the  
		commutation relation 
		\begin{equation}
			\Lieb{\hat x} {\hat p} = \i \hbar  \, \, . 
			\label{eq:commut}
		\end{equation}
	\end{enumerate}
\end{Proposition}
\begin{Proof}
\begin{subequations}
	Since both $\hat x$ and $\hat p $ are self-adjoint, they are 
	bounded (cf. \cite{Hall}*{Cor. 9.9}). \\ 
	From \eqref{eq:commut}, we find that $\hat x, \hat p $ 
	are non-zero 
	and thus have non-zero (operator) norms 
	$\norm {\hat x} , \norm {\hat p}$. Since $\hat x$ is self-adjoint, 
	it is normal 
	and thus $\norm{{\hat {x}}^2} = {\norm {\hat x} }^2$. 
	Now one proves by 
	induction that for all $n \in \N$ we have 
	\begin{equation}
		\Lieb{\hat x^n}{\hat p} = 
		\i n \hbar \, \hat x^{n-1}  \, . 
	\end{equation}
	Taking norms and applying the triangle inequality, we get 
	\begin{equation}
		n \hbar \leq 2 \norm {\hat x} \norm {\hat p } \, , 
	\end{equation}
	thus $\hat x$, $\hat p$ or both, are unbounded. 
	This is a contradiction. 
\end{subequations}
\end{Proof}
Therefore, even within the application of the $1$-particle 
Schr\"odinger theory, the Hilbert space formalism is 
inadequate. We refer to the book by Hall \cite{Hall} 
for alternative descriptions. 
\par 
Still, we would like to have a closer look at the 
expectation values of $\hat{x}^i$, $\hat{p}_j$,  
$\hat{L}_k$ and $\hat E$ in the context of the Madelung picture. 
Indeed, we will find that 
the Madelung picture gives a natural explanation for why 
the operators yield the physically correct 
expectation values - within Kolmogorovian probability theory 
(cf. \cites{Kolmogorov, Klenke}). 
In addition, the Madelung picture offers a natural, more intuitive 
formalism and, by making the analogy to Newtonian mechanics explicit, 
shows directly which observables are `physical'. 
\par 
In order to show this, we need to make some assumptions on the 
`regularity' of the involved functions and spaces:  
So given a Newtonian spacetime 
$\left( \mathcal Q, \d \tau, \delta, \mathcal O \right)$, 
we would like the operators $\hat{x}^i$, $\hat{p}_j$,  
$\hat{L}_k$ and $\hat E$ to be well-defined and satisfy 
\begin{equation}
	\inp{\Psi_t }{\hat A \Psi_t } = 
	\inp{\hat A \Psi_t }{\Psi_t } 
	\label{eq:phermit} 
\end{equation}
for each $t \in I$ and all `wave functions' 
\begin{equation}
	\Psi \colon \mathcal Q \to \,  \C 
	 \colon \left( t, \vec x \right) 
	 \to \Psi_t \left( \vec x \right) \, . 
\end{equation}
Observe that \eqref{eq:phermit} only makes sense for $\hat A = 
\hat E$, if we choose a potential 
$V\colon \mathcal Q \to \R$ and, for given $t$,  
interpret $\hat E$ as the Schr\"odinger operator 
\begin{equation}
 	-  \frac{\hbar^2}{2m} \Delta + 
	V \left( t, . \right) \, \, .
\end{equation}
Hence, from the Cauchy-Schwarz inequality and 
integration by parts, we conclude that $\Psi$ needs to be an 
element of%
\footnote{`$\Psi_t$ vanishes on $\partial \Omega_t$' means 
that for any sequence $(\vec{x}_n)_{n \in \N}$ in $\Omega_t$ 
converging to $\vec x \in \partial \Omega_t \subset \R^3$, we have 
$\lim\limits_{n \to \infty} \Psi_t \left( \vec x_n\right) = 0$.}  
\begin{equation}
	\begin{split}     
	\mathcal W \left( \mathcal Q, \C \right) :=   
	\biggl\lbrace 
	\Psi \in C^2 \left(\mathcal Q, \C \right)
	\biggm\vert \forall \, i \in \lbrace 1, 2, 3\rbrace\, 
	\forall t \in I
	\colon    \, \, \Psi_t, 
	x^i \Psi_t , \, \partd{\Psi_t}{x^i} , \, 
	\partd{\Psi_t}{t} \\ 
	\text{lie in} \, \,  
	L^2 \left( \Omega_t, \C \right) \, \,
	\text{and} \,  \Psi_t \, \, \text{vanishes on the boundary} \, \,
	\partial \Omega_t \, \, \text{in} \, \, \R^3    
	\biggr\rbrace 
	 \, , 
	\end{split}  
\end{equation}  
needs to satisfy the Schr\"odinger equation 
and that $\Omega_t$ needs to split into a product 
of three open intervals (up to a set of measure zero). 
The necessity of this unnatural assumption on 
$\Omega_t$ may be considered another 
indicator that the standard formulation of quantum 
mechanics is problematic. Commonly, one makes the implicit, 
stronger assumptions that each $\Omega_t$ is 
$\R^3$ (up to a set of measure zero), that $\Psi$ is 
a smooth solution of the Schr\"odinger equation 
with $\partd{\Psi_t}{t} \in L^2 \left( \Omega_t, \C \right)$,  
and that for each $t \in I$ the function $\Psi_t$ is an element 
of the \emph{space of ($\C$-valued) Schwartz functions} 
\begin{equation}
	\mathcal S\left( \Omega_t, \C \right)
	:= \bigl\lbrace \Psi_t \colon \Omega_t \to \C
	\bigm\vert  
	\forall \,\text{multi-indices}\, \alpha, \beta \colon 
	\left( x^\alpha \, \partial^\beta \, \Psi_t  \right) 
	\in L^2 \left( \Omega_t, \C \right) \bigr\rbrace 
\end{equation}     
\emph{on $\Omega_t$} 
(cf. \citelist{\cite{Becnel} \cite{Stein}*{\S 5.1.3 \& \S 6.2}}). 
For convenience, we choose the stronger assumptions 
on $\Psi$ and its domain in the following. 
\par
In order to relate everything 
to the Madelung picture, assume we are also given 
functions $\varphi \in C^\infty \left( \mathcal Q, \R 
\right)$ and $R \in C^\infty \left( \mathcal Q,   
[0, \infty ) \right)$ such that $\Psi = R e^{- \i \varphi}$. 
Since $\Psi_t$ is 
$L^2$-integrable for every $t$, 
we may normalize it to have unit $L^2$-norm. Then $\rho := R^2$, pulled 
back to $\Omega_t$, 
satisfies the mathematical axioms of a probability density 
(with respect to $\d^3 x$).  
\par
In the first instance, we may consider the position operators: 
\begin{equation} 
	\inp{\Psi_t}{\hat x^i \Psi_t}  = \int_{\Omega_t} \Psi^*_t 
	\left( \vec x \right) \, \left( \hat x^i \Psi_t \right) 
	\left( \vec x \right)
	\, \d^3 x 
	= \int_{\Omega_t}  x^i \, 
	\rho \left( t, \vec x \right) \d^3 x = \EXP{t, x^i} \, . 
\end{equation}
So we get the expectation value 
$\mathbb E \left( t, . \right)$ of the $i$th 
coordinate function with respect to the probability density 
$\rho (t, . )$ on $\Omega_t$. Then 
\begin{equation}
	\EXP{t, \vec x} :=  
	\left( \EXP{t, x^1},\EXP{t, x^2},\EXP{t, x^3} \right) 
	\in \R^3
\end{equation}
gives the mean 
position of the particle at 
time $t$ in $\Omega_t$. Moreover, in Kolmogorvian probability theory, 
we can ask for the expectation value of the 
position $\EXP{t, \vec x, U_t}$ on any other Borel 
set $U_t \in \mathcal B \left( \Omega_t \right)$, its standard deviation 
et cetera. 
This can also be done in the `von Neumann
philosophy' by multiplication with the indicator function  
\begin{equation}
		\chi_{U_t} \colon \Omega_t 
		\to \C \colon \vec x \to \chi_{U_t}
		\left( \vec x\right) := 
		\begin{cases}
			1 &, \vec x \in {U_t} \\
			0 &, \text{else}  
		\end{cases} \quad , 
		\end{equation}
but the product $\chi_{U_t} \, \Psi_t$ is usually not differentiable.  
\par
Second, consider the momentum operators: By \eqref{eq:phermit} 
for $\hat A = \hat{p}_i $, we have 
\begin{equation}
	\inp{\Psi_t}{\hat p _i \Psi_t} = \Re \inp{\Psi_t}{\hat p _i \Psi_t} \, 
\end{equation}
and thus 
\begin{align}
	\inp{\Psi_t}{\hat p _i \Psi_t} &= - \i \hbar \int_{\Omega_t} \Psi^*_t 
	\left( \vec x \right) \, \partd{\Psi_t}{x^i} 
	\left( \vec x \right)
	\, \d^3 x \\
	&= - \i \hbar  \int_{\Omega_t} \Psi^*_t 
	\left( \vec x \right) \, 
	\left( \partd{R}{x^i} \left( t, \vec x \right) 
	e^{-\i \varphi \left( t, \vec x \right)} - \i \partd{\varphi}{x^i} 
	\left( t, \vec x \right) \, \Psi_t \left( \vec x \right) \right) 
	\, \d^3 x \\
	&=\int_{\Omega_t} -\hbar \partd{\varphi}{x^i} 
	\left( t, \vec x \right) \, \rho \left( t, \vec x \right) 
	\, \d^3 x \\
	&= \EXP{t, m X^i} \, , 
\end{align}
using \eqref{eq:gradvarphi}. Therefore, if we are 
willing to interpret $m X^i$ as the
random variable for the $i$th component of the momentum, 
$\inp{\Psi_t}{\hat p _i \Psi_t}$ yields its expectation 
value. This interpretation is indeed a result of the 
correspondence principle (see section \ref{sec:interpretation}).   
In the linear operator formalism, however, we cannot 
simply replace the domain of the integral by some $U_t \in 
\mathcal B \left( \Omega_t \right)$, since $\hat p_i$ will no longer 
be interpretable as a momentum operator: The quantity   
\begin{equation}
	\int_{U_t} \d^3 x \, \, 
	\Psi_t^*\left( \vec x \right) \,  \left( \hat{p}_i \,    
	\Psi_t \right) \left( \vec x \right) 
\end{equation}
is usually not real. Hence 
the expectation value of the $i$th momentum in the region $U_t$ is 
not clearly defined in the `von Neumann philosophy'.  
Contrarily, in Kolmogorovian probability theory 
we only need to compute 
\begin{equation}
	\EXP{t, m X^i, U_t } := \int_{U_t} m X^i \left( t, \vec x \right) 
	\, \rho \left( t, \vec x \right)  \d ^3 x 
\end{equation}
to get the expectation value. 
\par
Concerning the energy operator, we need to exclude the zeros of 
$\Psi$ from $\mathcal Q$ to define the \emph{energy} $E$ via 
\begin{equation}
	E 
	:= \frac{m}{2} \vec X ^2 + V 
		-  \frac{\hbar^2}{2m} 
		\frac{\Delta \sqrt \rho}{\sqrt \rho} \, . 
\end{equation}
Note that $L^2 \left( \Omega_t, \C \setminus \lbrace 
0 \rbrace \right)$ is not a vector space, so 
`superposition' of wave functions requires a formal 
change of domain.   
By \eqref{eq:phermit} for $\hat A = \hat E$ and the 
Bernoulli-Madelung equation \eqref{eq:H}, an argument analogous 
to the one for the momentum operators indeed yields 
\begin{align}
	\inp{\Psi_t}{\hat E \Psi_t} &=  \i \hbar \int_{\Omega_t} \Psi^*_t 
	\left( \vec x \right) \, \partd{\Psi_t}{t} 
	\left( \vec x \right)
	\, \d^3 x \\
	&=\int_{\Omega_t}  \hbar \partd{\varphi}{t}
	\left(t, \vec x \right) \, \rho \left( t, \vec x \right) 
	\, \d^3 x \\
	&=\int_{\Omega_t} E \left(t, \vec x \right) \, 
	\rho \left( t, \vec x \right) \, \d^3 x \\
	&= \EXP{t, E} \, .    
\end{align}
\par
For the angular momentum operators the previous arguments 
can be repeated and one also finds the correct expectation value (cf. 
\cite{Holland}*{\S 3.8.2}).   
\par 
Therefore, our treatment not only suggests the inadequacy of the 
von Neumann approach, but also leads us to the 
following postulate.  
\begin{Postulate}
	\label{Post:Kolmogorov}
	Quantum theory is correctly axiomatized by Kolmogorovian 
	probability theory. 
\end{Postulate}   
Clearly, the statement is in potential conflict with von 
Neumann's projection postulate. Historically, von Neumann laid 
the mathematical foundations of modern quantum mechanics in 1932 
\cite{Neumann}, while Kolmogorov's axiomatization of modern 
probability theory \cite{Kolmogorov}
was published in 1933. 
Therefore von Neumann did not know of Kolmogorov's work at the time 
and that certain formulations of the Born rule \cite{Born} were in 
potential conflict with it. The view, commonly taken today with respect 
to this issue \cites{Streater, Redei}, is that quantum theory employs 
a more general notion of probability: There is a 
non-commutative probability theory and the Kolmogorovian 
approach is the particular, commutative case. 
However, apart from quantum theory, we are 
not aware of any applications of said generalization. Taking 
the historical context into account, 
it appears plausible that the projection postulate is wrong. In fact, 
the current view implicitly suggests that Kolmogorov has failed in 
axiomatizing probability theory in its most general framework, a 
statement we find troublesome. 
\par
Let us consider a specific example to make the potential 
conflict between the two approaches 
more explicit: If we ask 
for the probability of the particle in the state 
$\left( \rho, X \right)$ to have 
an energy $E \left(t, . \right)$ in the range 
$J \in \mathcal B \left(\R \right)$ at time $t \in I$, then, 
following Kolmogorov, this is given by 
\begin{equation}
	\int_{U_t} \iota^*_t\rho \, \d^3 x \quad , \, U_t 
	:= \left( E \left(t, . \right) \right)^{-1} \left( J \right)
	\subseteq{\Omega_t} \, . 
		\label{eq:Kolmo}
\end{equation}
In contrast, if $\hat E$, considered as a linear map 
from a linear subspace of 
$\mathcal S\left( \Omega_t, \C \right)$ to 
$L^2 \left( \Omega_t, \C \right)$, has a point spectrum 
$\set{E_n \in \R}{n \in \N}$ (cf. \cite{Hall}*{\S 9.4}) 
with mutually orthonormal  
eigenvectors $\left(\Phi_{n,t}\right)_{n \in \N}$, 
then, following von Neumann,  
the same probability is given by 
\begin{equation}
	\sum_{n \in \N, E_n \in J} \abs{\inp{\Phi_{n,t}}{\Psi_t}}^2 \, .
	\label{eq:Neumann} 
\end{equation}
It is clear that for $\Psi_t= \Phi_{n,t}$ both expressions yield 
either $1$ or $0$ for $E_n \in J$ or  $E_n \notin J$, respectively. 
For more general 
states the two do not appear to coincide, but this statement 
requires a proof (in terms of a counterexample) and maybe 
there exists an approximation.  
If the expressions \eqref{eq:Kolmo} and \eqref{eq:Neumann} differ, it is 
an empirical question which one of the two, if any, is correct. As a 
difference can only arise for time-dependent states, one should 
postpone this question until 
the corresponding relativistic theory has been laid out (see
\thref{Rem:time}). 

\section{Interpreting the Madelung Equations}
\label{sec:interpretation}	

As claimed previously, the Madelung equations are easier 
to interpret than the Schr\"odinger equation and it is the 
aim of this section to convince the reader 
of the truth of this statement. We first give a 
probabilistic, mathematical 
interpretation in section \ref{sec:interpretation1} and then proceed 
with a more speculative discussion in section \ref{sec:interpretation2}
on page \pageref{sec:interpretation2}.  
	
	\subsection{Mathematical Interpretation}
	\label{sec:interpretation1}
	
	Contrary to Madelung's interpretation of $\rho$ as  
	a mass density \cite{Madelung}, quantum mechanics is now 
	widely acknowledged 
	to be a probabilistic theory with $\rho$ being 
	the probability density 
	for 
	finding the particle within a certain region of space. This 
	is referred to 
	as the Born interpretation or ensemble interpretation, named after 
	Max Born \cite{Born}. 
	For a discussion on why other interpretations are not admissible, 
	we refer to
	\cite{Ballentine}*{\S 4.2} and, of course, Born's original article 
	\cite{Born}.  
	Taking this point of view, it is potentially fallacious to assume 
	that $X$ describes the actual velocity of the particle, as this 
	appears to oppose the probabilistic nature of the theory. 
	However, we can interpret $\vec j := \rho \vec X $ 
	as the probability current density, since then the continuity 
	equation \eqref{eq:continuity} reads 
	\begin{equation}
		0 = \partd{\rho}{t} + \nabla \cdot \vec j   \, .
	\end{equation}
	The physical meaning of this equation becomes more apparent when it 
	is formulated in the language of integrals:  Let 
	$N= N_0 \subseteq \Omega_0$ be an open set, $\Phi$ be the flow 
	of $X$ and assume 
	$N_t := \vec \Phi_t \left( N\right)\subseteq \Omega_t$ 
	exists for each $t \in I$. This is for instance the case, if 
	we have the situation of \thref{Prop:topOQ}. Then 
	the Reynold's transport theorem \cite{Acheson}*{\S 6.3} implies 
	that for such an $N$ moving along the flow 
	\begin{equation}
		\partd{}{t} \int_{N_t} \iota^*_t\rho \,  \d^3 x = 0 \,  
		\label{eq:rhoconserved}
	\end{equation}
	for all times $t \in I$ - provided $\iota^*_t \rho$ is integrable
	on $\Omega_t$ for all $t \in I$.%
	\footnote{On a technical note, to assure convergence of 
	\eqref{eq:rhoconserved}, we also require that the function 
	\begin{equation*}
	N \to \R \colon
	\vec x \to
	\sup_{t \in I} \abs{\left( 
	\partd{\rho}{t} \bigl(t, \vec \Phi_t \left( \vec x\right) \bigr)  + 
	\left( \nabla \cdot \left( \rho \vec X \right)  \right)  
	\bigl(t, \vec \Phi_t \left( \vec x\right) \bigr) 
	\right) 
	\,
	\det \biggl( \biggl(\partd{ \vec \Phi_t}{\vec x} \biggr) 
	\left( \vec x \right) \biggr)  
	}
	\end{equation*}
	is bounded and integrable over $N$. This is trivially true, if the 
	continuity equation holds. 
	} Integrability is assured by the fact that $\iota^*_t\rho$ 
	is a probability 
	density for all $t \in I$ and, following the discussion in section 
	\ref{sec:operator}, one may even assume that it is Schwartz, 
	i.e. 
	\begin{equation}
		\iota^*_t \rho \in \mathcal S \left(\Omega_t, \R_+ \right) 
		\quad \quad
		\forall t \in I \, .
	\end{equation}
	By Gau{\ss}' divergence theorem we have
	\begin{equation}
		\int_{\partial N_t} \vec j_{\iota_t} \cdot \d \vec A_t = 
	 	- \int_{N_t} \iota^*_t\partd{\rho}{t} \, \, \d^3 x 
	 \, .
	 \label{eq:flux}
	\end{equation}
	Equation \eqref{eq:rhoconserved} states, that the probability that 
	the particle is found within $N$ stays conserved, if $N$ moves 
	along the flow of $X$. \eqref{eq:flux} states that the probability 
	flux leaving $N_t$ is the probability current through its 
	surface obtained from $\vec j$. 
	\begin{figure}[h]  
	\centering
	\includegraphics[scale=0.25]{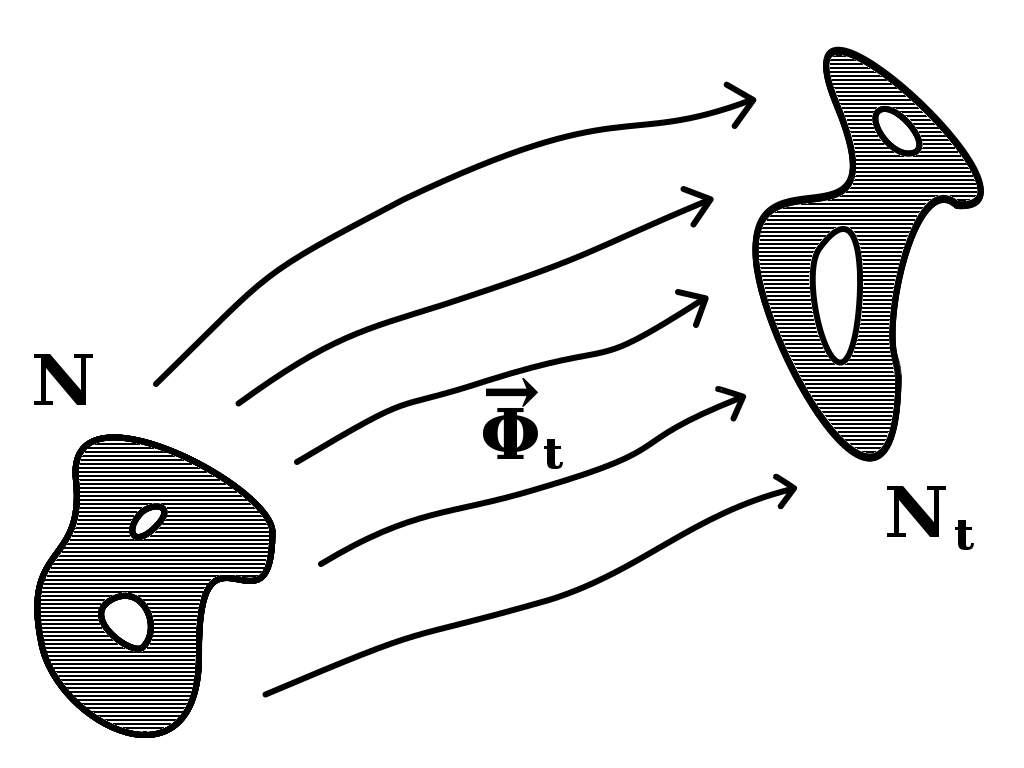}
	\caption{The region $N$, where the particle is located with 
	initial probability $ \mathbb P \left(0, N\right) =
	\int_N \iota_0^* \rho \, \d^3x$, propagates along the flow 
	$\vec \Phi$ in space. After time $t>0$, the region has 
	been transformed to 
	$N_t = \vec \Phi_t \left( N\right)$. The probability to find the 
	particle, as well as the type and number of holes within the region, 
	stays conserved, but the region may be distorted, shrunk or 
	expanded.}  
	\label{fig:propagation}
	\end{figure}%
	We conclude that the primary importance of the drift field lies 
	in the fact 
	that its flow describes the probabilistic propagation of the system. 
	If, for example, we take $N$ to be a ``small'' region with 
	$95 \%$ chance of finding the particle and we let this region 
	``propagate'' along the drift flow, then this probability will 
	not change over time. 
	However, it might happen that the volume of $N$ increases
	or decreases. 
	Under appropriate assumptions on 
	convergence, 
	the change of volume of $N$ is given by
	\begin{equation}
		\partd{}{t} \int_{N_t} \d^3 x = \int_{N_t} \bigl( \nabla \cdot 
		\vec X\bigr)_{\iota_t} \, \d^3x \, ,
	\end{equation}
	again by the Reynold's transport theorem. Therefore, 
	the divergence of 
	the spacelike component of the drift field is a measure of how $N$ 
	spreads or shrinks with time. Moreover, ``holes'' in $\Omega$, 
	appearing for instance due to the vanishing of $\rho$, can also be 
	viewed as propagating with time (see \thref{Prop:topOQ} on page 
	\pageref{Prop:topOQ} 
	and \cite{Holland1}*{p. 11}), due to the fact that the 
	`spacelike part' of
	the drift flow $\vec \Phi_t$ is a diffeomorphism for each time $t$.
	The situation is schematically depicted in figure 
	\ref{fig:propagation}.
	\begin{Remark}[Particle structure]
	\label{Rem:sparticles}
		In the Madelung picture particles are treated as 
		(approximately) point-like, 
		since the support of $\iota^*_t\rho$ can be made 
		arbitrarily small. 
		In this context, we would also like to remark that, if 
		the initial probability 
		density is given by a Gau{\ss}ian with standard deviation 
		$\sigma \in \R_+$ and the initial drift field is constant, 
		then solving 
		the Madelung equations and taking the limit $\sigma \to 0$ 
		might make it possible to assign trajectories, energies, etc. to 
		individual particles.
	\end{Remark}
	\par 
	Yet this discussion does not fully answer the question how the 
	drift field itself is to be interpreted and 
	practically	determined. The following result, central to 
	the resolution of
	this question, was conjectured by Christof Tinnes (TU Berlin) and a 
	weaker version had already been discovered by P. 
	Ehrenfest \cite{Ehrenfest}. 
	\begin{Theorem}[Expectation value of the drift field]
	\label{Thm:interpretX}
	\begin{subequations}
	Let $\left( \mathcal Q , \d \tau, \delta, 
	\mathcal O \right)$ be a Newtonian spacetime, let $\Omega_t$, $I$ be 
	defined as in 
	\eqref{eq:Omega}, $X \in \mathfrak X _{Nt} 
	\left( \mathcal Q \right)$ be a Newtonian observer vector field 
	with flow $\Phi$, 
	$\rho \in C^\infty \left( \mathcal Q , \R_+ \cup \lbrace 0 \rbrace 
	\right)$ a positive, real function such that 
	$\iota_t^* \rho$ is Schwartz and a probability density for 
	all $t \in I$, 
	and assume the continuity equation 
	\eqref{eq:continuity} holds. 
	Define for all $t \in I$, 
	$U_t \in \mathcal B \left( \Omega_t \right)$ and $f \in  
	C^\infty \left( \mathcal Q, \R \right)$ the 
	\emph{expectation value of $f$ at time $t$ over $U_t$}: 
	\begin{equation}
		\mathbb E \left(t, f, U_t \right) := 
		\int_{U_t} \iota^*_t f \, \, \iota^*_t\rho\,  \d ^3 x \, .
		\label{eq:DefE}
	\end{equation}
	Then for every $N \in \mathcal B \left( \Omega_0 \right)$, such that
	the functions 
	\begin{equation}  
		N \to  \R \colon 
		\vec x \to \sup_{t \in I} \abs{ X^i 
		 \left( \Phi_t \left( 0, \vec x \right)\right) \, 
		 \rho 
		 \left( \Phi_t \left( 0, \vec x \right)\right) \, 
		 \det \biggl( \biggl(\partd{ \vec \Phi_t}{\vec x} \biggr) 
		\left( \vec x \right) \biggr)} 
		\label{eq:convergenceE}
	\end{equation}
	are bounded and integrable for each $i \in \lbrace 1, 2, 3 \rbrace$,
	and every 
	$t \in I$ s.t. $N_t := \vec \Phi_t \left( N \right)$ exists, we have  
	\begin{equation}
			 \mathbb E \left(t,  \vec X, N_t \right) = \frac{\d}{\d t} 
			 \mathbb E \left(t,  \vec x, N_t \right) \, .
			 \label{eq:interpretX}
	\end{equation}
	Here we defined 
	\begin{equation}
	\mathbb E \left(t, \vec X, U \right) := \left( \mathbb E 
	\left(t, X^1, 
	U \right),\mathbb E \left(t, X^2, U \right), 
	\mathbb E \left(t, X^3, U \right)\right) \in \R^3 \, . 
	\end{equation}
	\end{subequations}
	\end{Theorem}
	\begin{Proof}
	\begin{subequations}
	The theorem is a corollary of the Reynold's transport 
	theorem, formulated as in
	\cite{Chorin}*{p. 10}. Since $\vec \Phi_t$, if defined, 
	is a homeomorphism onto 
	its image, the respective topologies coincide, and hence 
	$N$ is a Borel set if and only if $N_t$ is a Borel set.
	We now note that for 
	$f \in C^\infty \left( \mathcal Q, \R \right)$ with 
	\begin{equation}  
		N \to  \R \colon 
		\vec x \to \sup_{t \in I} \abs{ X_{\Phi_t \left( 0, 
		\vec x \right)} 
		\left( f \right) \, 
		 \rho 
		 \left( \Phi_t \left( 0, \vec x \right)\right) \, 
		 \det \biggl( \biggl(\partd{ \vec \Phi_t}{\vec x} \biggr) 
		\left( \vec x \right) \biggr)} 
	\end{equation}
	bounded and integrable and assuming convergence of the respective 
	integrals, the 
	continuity equation \eqref{eq:continuity} implies 
	\begin{align}
		\frac{\d}{\d t} \int_{N_t} 
		\iota_t^* \left( 
		f \rho  \right) \, \d^3 x &=  \int_{N_t} \left( 
		\partd{\left(f \rho \right)}{t} + \nabla \cdot 
		\left( f \rho \vec X \right) \right)_{\iota_t} \d^3 x \\
		&=  \int_{N_t} \left( 
		\partd{f}{t} + \nabla  f  \cdot \vec X   
		\right)_{\iota_t} \, \iota_t^*\rho \, \d^3 x \\
		&= \int_{N_t}  X_{\iota_t} \negthinspace \left( f 
		\right) \, \iota_t^*\rho \, \d^3 x  \,  .
	\end{align}
	Now set $f= x^i$ with
	$i \in \lbrace 1,2, 3\rbrace$, observe that 
	$\iota^*_t \left( x^i \rho \right)$ is integrable 
	and due to $X\left(x^i \right)
	= X^i$ the result follows. 
	\end{subequations} 
	\end{Proof} 
	Equation \eqref{eq:interpretX} roughly means that the expectation 
	value of the drift field in some region $N$ moving along its flow 
	is given by the velocity of the expectation value of the 
	position in $N_t$. 
	Moreover, \thref{Thm:interpretX} can be used to find an even 
	more direct interpretation of the drift field. 
\begin{Corollary}[Interpretation of the drift field]
\label{Cor:interpretX}
\begin{subequations}
	Let $\left( \mathcal Q , \d \tau, \delta, 
	\mathcal O \right)$ be a Newtonian spacetime, let $\Omega_t$, $I$ be 
	defined as in 
	\eqref{eq:Omega}, $X \in \mathfrak X _{Nt} 
	\left( \mathcal Q \right)$ be a Newtonian observer vector field 
	with flow $\Phi$, 
	$\rho \in C^\infty \left( \mathcal Q , \R_+ \rbrace 
	\right)$ a strictly positive, real function such that 
	$\iota_t^* \rho$ is Schwartz and a probability density for all 
	$t \in I$,
	and assume the continuity equation 
	\eqref{eq:continuity} holds. \\
	Define $\mathbb E$ as in \thref{Thm:interpretX} and 
	for $t \in I$, $U_t \in 
	\mathcal B \left( \Omega_t \right)$ 
	let $\mathbb P \left(t, U_t \right) 
	:= \mathbb E \left(t, 1, U_t \right)$  be the probability of $U_t$. 
	Further, for $\epsilon \in \R_+$, $\vec y \in \Omega := \Omega_0$ 
	define 
	\begin{equation}
		N^\epsilon \left( \vec y \right) := \set{\vec x \in \Omega }{ 
		\dist\left( \vec x , \vec y \right) < \epsilon}
	\end{equation}
	with $\dist$ denoting the Riemannian distance on 
	$\left( \Omega, \iota^*_0 \delta \right)$. \\ 
	Then for every $\epsilon \in \R_+$, $\vec y \in \Omega$ and every 
	$t \in \R$ such that $N_t^\epsilon \left( \vec y \right) := 
	\vec
	\Phi_t \left( N^\epsilon \left( \vec y \right) \right) 
	\subseteq \Omega_t$ 
	exists and the functions \eqref{eq:convergenceE} 
	for $N =  N^\epsilon \left( \vec y \right)$,  
	$i \in \lbrace 1, 2, 3\rbrace$
	are bounded and integrable, we have 
	\begin{equation} 
		\vec X_{\Phi_t\left(0,\vec y \right)} = \lim_{\epsilon \to 0} 
		\frac{\frac{\d}{\d t} 
			 \mathbb E \left(t, \vec x, N_t^\epsilon 
			 \left( \vec y \right)\right)}
			 {\mathbb P \left(t, 
			 N_t^\epsilon \left( \vec y \right) \right)}
		\, .
	\end{equation}
\end{subequations}
\end{Corollary}
\begin{Proof}
\begin{subequations}
	 By assumption $N_t^\epsilon \left( \vec y \right)$ is defined, 
	 thus the 
	 restriction $\vec \xi_t$ of 
	 $\vec \Phi_t$ to the open submanifold 
	 $N^\epsilon \left( \vec y \right)$ and its image is also 
	 defined. Since 
	 $N^\epsilon \left( \vec y \right)$ is open in $\Omega$, it is 
	 Borel-measurable. 
	 As $N_t^\epsilon \left( \vec y \right)$ is also open, 
	 non-empty and $\rho >0$, it follows 
	 $\mathbb P \left(t, N_t^\epsilon \left( \vec y \right) \right) > 0$. 
	 By 
	 \thref{Thm:interpretX} we have 
		\begin{equation}
		\frac{\frac{\d}{\d t} 
			 \mathbb E \left(t, \vec x, N_t^\epsilon 
			 \left( \vec y \right)\right)}
			 {\mathbb P \left( t, 
			 N_t^\epsilon \left( \vec y \right) \right)} 
			 = \frac{
			 \mathbb E \left(t, \vec X, N_t^\epsilon 
			 \left( \vec y \right)\right)}
			 {\mathbb P \left( t, 
			 N_t^\epsilon \left( \vec y \right) \right)} 
			 =  \frac{ \int_{ N_t^\epsilon \left( \vec y \right)} 
			 \iota^*_t \negthinspace \left( 
			 X^i \, \rho \right) \, \d^3 x }{
			 \int_{ N_t^\epsilon \left( \vec y \right) }
			 \iota^*_t\rho \, \d^3 x } \, e_i    
			 \, ,
		\end{equation}
	 where $e_i$ is the coordinate basis vector for 
	 $i \in \lbrace 1,2,3\rbrace$. For 
	 every $\epsilon' \in \R$ with $0 < 
	 \epsilon' \leq \epsilon$ the point 
	 $\vec \Phi_t\left( \vec y \right)$ is in $N_t^{\epsilon'} 
	 \left( \vec y \right)$ 
	 by definition. Moreover, the diameter of $N_t^{\epsilon'} 
	 \left( \vec y \right)$ tends to 
	 zero as $\epsilon' \to 0$ due to continuity of 
	 $\vec \xi_t$. Considering 
	 $\iota^*_t\rho \, \d^3 x$ as a volume form and applying 
	 \cite{Fleming}*{\S 8.4, Lem. 1} 
	 yields 
		\begin{equation}
			X^i \circ \Phi_t\left(0,\vec y \right) \, e_i = 
			\lim_{\epsilon \to 0}
			 \frac{ \int_{ N_t^\epsilon \left( \vec y \right)} 
			 \iota^*_t \negthinspace \left( 
			 X^i \, \rho \right) \, \d^3 x }{
			 \int_{ N_t^\epsilon \left( \vec y \right) }
			 \iota^*_t\rho \, \d^3 x } \, e_i  \, .
			 \label{eq:actualX}
		\end{equation}
	 Identifying $e_i$ with $\partial_i$, such that we may write 
	 $\vec X = X^i e_i$, completes the proof. 
	\end{subequations}
	\end{Proof}
	\thref{Cor:interpretX} yields a direct interpretation of the 
	drift field in terms
	of probabilistic quantities. Since 
	$\mathbb P \left(t, 
	N_t^\epsilon \left( \vec y \right) \right)$ is 
	the probability of 
	the particle to be found in the set $N_t^\epsilon 
	\left( \vec y \right)$, equation
	\eqref{eq:actualX} states that the drift field gives 
	the infinitesimal velocity of the expectation value of the 
	particle's position per unit 
	probability of finding the particle in this region. That is, 
	if the particle is certain 
	to be found in a small enough region of space, the 
	(approximately constant) drift 
	field gives the velocity of the expectation value of the 
	particle's position 
	in this region. To be able to 
	make practical use of this statement, we postulate the following. 
	\begin{Postulate}[Interpretation of the drift field]
	\label{Post:interpretX}
	\begin{subequations}
		The velocity of the expectation value of the position 
		for an ensemble of particles in 
		a small region of space is equal to the average velocity 
		of the ensemble of 
		particles in that region. 
	\end{subequations} 
	\end{Postulate}
	\thref{Post:interpretX} states that one can determine the drift 
	field at each point 
	by determining the average velocity of the particles hitting 
	the point. 
	Within a stochastic analogue of the theory, it should be possible 
	to assign a precise mathematical meaning to 
	\thref{Post:interpretX} and determine 
	its truth value, but for our purposes here we shall assume the 
	truth of the statement without proof.%
	\footnote{It should be noted that there have already been 
	attempts to find 
	a stochastic formulation of the Madelung equations within 
	the so-called 
	theory of 'stochastic mechanics', developed mainly by E. 
	Nelson. See e.g. 
	\cite{Nelson3} for a review.} 
	In this context, 
	it is useful to observe that, by construction, the domains 
	$\dom \rho$ and $\dom X$ 
	are equal and hence the drift field only needs to be given 
	where particles can 
	actually be found. 
	This compatibility of the interpretation of the drift field with 
	the ensemble/Born interpretation of quantum mechanics is also the 
	point where the Madelung picture differs \cite{Tsekov1} from the 
	Bohmian interpretation \cites{Bohm1,Bohm2}. Indeed, our discussion 
	shows how the ensemble interpretation naturally coheres with the 
	mathematical formalism of the Madelung picture, once the 
	Born rule is assumed. 
	\par
	We are now in a position to practically apply the formalism. 
	This is implicitly 
	related to the question whether the wave function is ``objective'' or 
	``an element of physical reality'' \cite{Einstein3}. We translate this
	as being measurable in the physical sense. 
	In the Madelung picture, this amounts to the question whether the 
	probability density and the drift field are measurable, both of which 
	are probabilistic quantities. 
	\par 
	Consider now, for example, 
	a particle gun that is used in the set-up of an arbitrary quantum 
	mechanical experiment, in principle describable via the 
	Madelung equations.  
	Before we run the experiment, we need to collect 
	initial data to solve 
	the Madelung equations. According to the Born interpretation, we 
	do this 
	by placing a suitable detector in front of 
	the particle gun and measuring the 3-dimensional (!)
	distribution of positions 
	(where the particle hits) and, following \thref{Cor:interpretX} and 
	\thref{Post:interpretX}, the average momenta (how hard the 
	particle hits) at each point. If we run the experiment 
	infinitely often, which is of course 
	an idealization, we expect to obtain
	a smooth probability density $\rho$ and a smooth drift momentum field 
	$P = m X$ in space at time $t=0$. We can then run the 
	actual experiment (ideally) infinitely 
	often and measure the distribution of positions and the 
	average momenta at each position. 
	If the Madelung equations provide a correct description of the 
	physical process and the detectors are ideal, this data will 
	coincide with the 
	one predicted by the Madelung equations for the given initial data. 
	Therefore, both 
	$\rho$ and $\vec X$ are measurable and thus objective as 
	probabilistic quantities. No measurement problem appears in this 
	case: The 
	time evolution of the probability density is deterministic and 
	the theory makes only probabilistic statements on 
	individual measurements. Furthermore, the mathematical 
	formalism makes no statement on the process 
	of measurement itself. 
	\begin{Remark}[Ideal detectors \& the Heisenberg Relation]
		\label{Rem:idealdetectors}
		Within the Copenhagen interpretation of the Schr\"odinger 
		theory, it is possible 
		to deny the existence of ideal detectors on the basis of the 
		(here one-dimensional) Heisenberg inequality
			\begin{equation}
			\Delta x \, \Delta p \geq \frac{\hbar}{2} \, .
			\label{eq:Heisenbergun}
			\end{equation}
		However, if one employs the ensemble interpretation and 
		observes that 
		\eqref{eq:Heisenbergun} is derivable within the 
		Schr\"odinger picture, one is forced 
		to conclude that \eqref{eq:Heisenbergun} is not a statement 
		on individual particles,   
		but one of statistical nature. That is, within the 
		ensemble interpretation,  
		\eqref{eq:Heisenbergun} does not support the interpretation it 
		is given within the 
		Copenhagen point of view, which itself has been subject 
		to criticism for a long time
		\cite{Einstein3}. In fact, the Heisenberg inequality is a 
		general statement  
		on Fourier transforms \cite{Stein}*{Thm. 4.1} and $\Delta p$ 
		is not the 
		standard deviation for the momentum given by the 
		Madelung picture in 
		conjunction with the Kolmogorovian probability 
		theory (see section 
		\ref{sec:operator} on page \pageref{sec:operator}) and 
		\cite{Holland}*{\S 6.7.3 \& \S 8.5}. Hence, if 
		\thref{Post:Kolmogorov} is adapted and 
		$\Delta p$ stands for the standard deviation in momentum, the 
		Heisenberg inequality is incorrect. 
		We conclude that the Heisenberg inequality does not put 
		any restrictions on 
		the precision of individual measurements and it does not appear 
		to bear any physical significance within the Madelung picture. 
	\end{Remark}
	For further discussion on the interpretation of 
	quantum mechanical states, we again refer to 
	\cite{Ballentine}*{\S 9.3}.
	\par 
	Having concluded our discussion on the continuity equation 
	\eqref{eq:continuity}, we now interpret the irrotationality of the 
	drift field \eqref{eq:irrotX}. Equation \eqref{eq:irrotX} has 
	a direct interpretation using the fluid dynamics analogue, namely 
	that it has vanishing vorticity   
	\begin{equation}
		\vec \sigma := \nabla \cross \vec X  \, .
	\end{equation}
	Following \cite{Acheson}*{\S 1.4}, half of the vorticity 
	``represents the 
	average angular velocity of two short fluid line elements that 
	happen, at that instant, to be mutually perpendicular''. 
	This statement derives itself from 
	\cite{Chorin}*{Eq. 2.1}. Returning to the situation as depicted 
	in figure 
	\ref{fig:propagation} on page \pageref{fig:propagation}, we therefore 
	find that the irrotationality of $X$ means that $N$ does not shear or 
	rotate when propagating along the flow of $X$. 
	Thus any distortion of the region $N$ over time 
	is due to shrinkage or expansion, not shear or rotation. 
	Moreover, the vorticity of the 
	velocity field of a fluid gives the infinitesimal circulation density, 
	which is derived from the integral definition of the curl operator 
	\cite{McKay}. In particular, if the vorticity of $X$ vanishes, 
	then 
	for all curves $\gamma$ in some $\Omega_t$ joining any two points in a 
	simply connected, open subset $U \subseteq \Omega_t$, 
	the value of
	\begin{equation}
		\int_\gamma \iota^*_t \left( \delta \cdot X \right)
	\end{equation} 
	depends only on the endpoints. 
	Several researchers \cites{Takabayasi1, Spera, Salesi} have  
	already suggested that quantum mechanical spin is 
	related, or even equivalent, to the vorticity of the drift field.
	Indeed, the factor of $1/2$ is 
	very suggestive and there exists already works in the 
	literature concerning 
	this question 
	\citelist{\cite{Bohm4} \cite{Bohm5} \cite{Hestenes1} 
	\cite{Holland}*{\S 9 \& \S 10}
	\cite{Janossy3} \cite{Takabayasi1} \cite{Salesi}}, 
	but as this article is only concerned with single-Schr\"odinger 
	particle systems, we do not discuss this relation here. 
	Moreover, we are not in 
	a position to pass judgement or elaborate on this relation yet. 
	For a 
	mathematical introduction to vorticity, see
	\cite{Chorin}*{\S 1.2} and for a very illustrative, freely accessible, 
	graphical exposition of the curl operator, see \cite{Nykamp}. We also 
	highly recommend watching the movie on vorticity \cite{NCFMF} 
	from the point of view advocated here.
	\par
	It remains to interpret the Newton-Madelung equation 
	\eqref{eq:Madelung}. 
	Due to the fact that the Newton-Madelung equation 
	\eqref{eq:Madelung} reduces 
	to Newton's second law \eqref{eq:Newton} for masses that are ``large'' 
	(as compared, e.g. to the 
	Planck mass), the classical limit of the entire model is quite easily 
	obtained by looking at the large mass approximation 
	of the Madelung equations: 
	\begin{subequations}
	\label{eq:pNewton} 
	\begin{align}
		m \dot X = \vec F \, ,   
		\\
		\nabla \cross \vec X = 0 \, ,\\
		\partd{\rho}{t} + \nabla \cdot \left( \rho \vec X \right) = 0 \, .
	\end{align}	
	\end{subequations}
	As the prior discussion also applies here, 
	these equations yield a probabilistic version of Newtonian 
	mechanics.%
	\footnote{This is another instance, where the von Neumann approach 
	to probability (see section \ref{sec:operator}) 
	leads to questionable results: Why should one change the 
	probability theory in the large mass-approximation?  
	} 
	This makes them compatible with the ensemble interpretation and 
	the requirement that 
	Newtonian mechanics must hold in some limit, as stated in the 
	introductory discussion. Note again that $\rho$ should not vanish 
	on its specified 
	domain and that $\dom \rho = \dom X$. 
	Hence, following \thref{Thm:equivalence}, we observe that the 
	first ad hoc modification 
	of Newtonian mechanics in quantum mechanics, i.e. the replacement 
	of Newton's $2$\textsuperscript{nd} law with the 
	Schr\"odinger equation, amounted to implicitly going over to 
	a probabilistic formalism and adding the Bohm force.  
	We are thus motivated 
	to postulate a new ``principle of classical correspondence'', 
	which was originally postulated by Niels Bohr in terms of 
	quantum numbers \cite{Bohr2}. 
	\begin{Postulate}[Non-quantum limit]
	\label{Post:nq-limit}
	For large masses, non-relativistic quantum theory, that is 
	quantum mechanics, 
	reduces to a probabilistic version of Newtonian Mechanics. 
	\end{Postulate}
	Experimentally, this limit can be made quantitative by 
	sending particles of 
	different mass through a double slit and finding the value $m_q$ 
	at which equations \eqref{eq:pNewton} cease to be a good description. 
	The so-called 
	classical limit is then $m/m_q \gg 1$, which is independent of 
	units. On a theoretical level, one could non-dimensionalize 
	the Madelung equations 
	and look at the magnitude of the perturbation introduced by the Bohm 
	force, but we abstain from doing this here. 
	\par 
	A generalized version of the Newtonian limit is also immediate.
	\begin{Postulate}[Generalized Newtonian limit]
	\label{Post:gN-limit}
	For large masses, small velocities and negligible spacetime 
	curvature, relativistic quantum theory 
	reduces to a probabilistic version of Newtonian Mechanics. 
	\end{Postulate}
	Clearly, it is only the Newton-Madelung equation that 
	changes under the non-quantum limit and our 
	previous discussion on the other two Madelung equations 
	remains valid in this case. An interpretation of the 
	Newton-Madelung equation thus has to focus
	on the Bohm force 
	\begin{equation}
		\vec F _ {B} \left( \rho \right) = 
		\frac{\hbar^2}{2m} \nabla 
		\frac{\Delta \sqrt{\rho}}{\sqrt{\rho}} \, .
		\label{eq:noise}
	\end{equation}
	A peculiar feature of this term, as well 
	as the Madelung equations as a whole, 
	is the invariance under the scaling transformation 
	$\rho \to \lambda \rho$ with $\lambda \in \R \setminus \lbrace 0
	\rbrace$. Hence the Madelung equations do not change, if $\rho$ is 
	not normalized, a fact that could be useful for the generalization to 
	multi-particle systems (see section \ref{sec:annihilation}). 
	For the interpretation of
	\eqref{eq:noise}, this means that the value 
	of the term is not influenced by the 
	value of the probability density, but only by its shape. 
	\par
	This property is to be expected a priori by the principle of 
	locality: If 
	we have two \emph{isolated ensembles} specified by the states 
	$\left( \rho_1, X_1 \right), \left( \rho_2, X_2 \right)$, 
	respectively, satisfying 
	$\dom \rho_1  \cap \dom \rho_2 = \emptyset$, then describing 
	them separately from 
	another via the Madelung equations or together should not make 
	any difference in terms of dynamics. 
	More precisely, for 
	$A \subseteq \mathcal Q := \dom \rho_1 \cup \dom \rho_2$
	we again define the indicator function 
	\begin{equation}
		\chi_A \colon \mathcal Q
		\to \R \colon x	\to \chi_A \left( x\right) := 
		\begin{cases}
			1 &, x \in A \\
			0 &, \text{else}  
		\end{cases}
		\end{equation}
	of $A$, 
	$\chi_1 := \chi_{\dom \rho_1}$, $\chi_2 := \chi_{\dom \rho_2}$, 
	and we set $\rho := \chi_1 \, \rho_1 /2 + \chi_2 \, \rho_2 /2$, 
	as well as  
	$X :=\chi_1 \, X_1 +\chi_2 \, X_2$. As for 
	$\dom \rho_1 \cap \dom \rho_2 
	= \emptyset$ both $\rho$ and $X$ are smooth, we can 
	now check whether they are a solution to the Madelung equations. 
	We indeed have 
	\begin{equation}
		\frac{\Delta \sqrt{\rho}}{\sqrt{\rho}} = 
		\frac{\Delta \sqrt{\rho_1}}{\sqrt{\rho_1}} \chi_1
		+  \frac{\Delta \sqrt{\rho_2}}{\sqrt{\rho_2}} \chi_2   \, ,
		\label{eq:isolatedsys}
	\end{equation}
	and the other two equations also separate, as required by 
	this consistency condition. 
	Interestingly, if the domains overlap and $\rho$ and $X$ 
	are sufficiently smooth,
	then \eqref{eq:isolatedsys} does not hold 
	and thus $\left( \rho, X\right)$, as defined, is 
	in general not a solution of the Madelung equations. This 
	can be explained by the fact 
	that one gets an entirely new ensemble in that case and hence 
	the non-linearity of 
	\eqref{eq:noise} in $\rho$ is not necessarily a defect of the 
	theory. Non-linearity 
	here means that $\vec F _B$ is not linear (and not even defined), 
	if extended 
	to the vector space $C^\infty \left( \mathcal Q, \R \right)$ 
	via \eqref{eq:noise}.  
	This point of view 
	potentially explains the results of the double slit experiment, 
	but the 
	statement remains
	of speculative nature unless a careful mathematical treatment 
	is given. 
	\par    
	As compared to the respective Newtonian theory, the term 
	\eqref{eq:noise} also causes an additional coupling between the 
	drift field and the 
	probability density,
	that goes beyond the requirement that the flow of
	the drift field is probability preserving in the sense of 
	the continuity equation \eqref{eq:rhoconserved}. 
	Thus how the probability density changes in space
	determines how the drift field behaves and vice versa in a 
	nonlinear manner. 
	Consequently, perhaps quite surprisingly to some, it is a nonlinearity 
	that causes much of quantum-mechanical behavior. 
	\par
	Intuitively, \eqref{eq:noise} represents a kind of noise 
	that disappears for large masses, which leads us propose an 
	alternative terminology for the term \eqref{eq:noise}: 
	\emph{Quantum noise} or \emph{Bohm noise}. 
	
	\subsection{Speculative Interpretation}
	\label{sec:interpretation2}
	
	At this point, we can only speculate on the origin of the (quantum) 
	noise term, but there is a particular interpretation that 
	suggests itself given our current knowledge of physics and 
	considering that the term is only relevant for small masses. 
	Before we proceed, we would like to stress that what 
	follows is speculative and should be considered 
	as standing fully separate from the rest of the article. 
	We understand the controversial nature of various attempts 
	of interpreting quantum mechanics \cite{Baggot}, but we 
	consider the need to find a coherent interpretation of 
	the equations as vital for the progress in the field. Needless 
	to say, any interpretation of a theory of nature has to 
	exhibit a strong link between 
	the applied theory and the mathematical formalism and may not 
	contradict either. In the following, we will speak about 
	quantum mechanics in general and not limit ourselves to the 
	$1$-particle Schr\"odinger theory.
	\par
	In 2005 Couder et al. \cite{Couder1} discovered that a silicon 
	droplet on the surface of a vertically oscillating silicon 
	bath remains stationary in a certain frequency regime, in which 
	coalescence is prevented. When the sinusoidal, vertical force on 
	the bath reaches a critical amplitude, the droplet  
	begins to accelerate and can be made to ``walk'' on the 
	surface of the bath \cite{Couder2}. Surprisingly, this basic setup 
	is a macrospscopic quantum analogue and can be used to build 
	more complicated ones. For a mathematical model see 
	\cite{Protiere}, and for a brief summary we refer to \cite{Bush}. 
	If two droplets approach each other, they 
	either scatter, coalesce or lock into orbit. In the latter case, 
	Couder et al. observed that the distances between the averaged orbits 
	is approximately one Faraday wavelength \cite{Couder2}, which 
	means that they are ``quantized'', in the sense of being 
	discrete. Moreover, when Couder and Fort studied the 
	statistical behavior of such a droplet passing
	a double-slit wall, it resembled the one found in the 
	quantum-mechanical analogue \cite{Couder3}. The fact that 
	Eddi et al. were 
	in addition able to establish the occurrence of tunneling for the 
	droplet \cite{Eddi}, suggests that a qualitatively similar 
	behavior occurs in the microscopic realm. How is this to be 
	explained? 
	\par 
	A physicist in the beginning of the twentieth century might 
	have justified this analogy via a vibration of the ether: 
	If the particle is massive enough, the
	influence of the ether's motion on the particle is negligible and it 
	behaves according to Newton's laws. Yet when the mass of the 
	particle is 
	small, the more or less random vibrations of the ether cannot be 
	neglected any more and a statistical description, that models the 
	noise caused by the ether's vibration via \eqref{eq:noise}, 
	becomes necessary. 
	\par 
	Of course, this explanation is flawed. 
	The Michelson-Morley 
	experiment famously ruled out any influence of the ether's motion
	on light \cite{Michelson} and an influence on matter had 
	not been observed, which ultimately led to 
	the creation of the theory of special relativity \cite{Einstein0}. 
	In addition, the existence of the ether would
	have established the existence of a preferred `rest frame', being 
	the one in which the ether is stationary, which in turn, if the 
	above interpretation were correct, would suggest a natural tendency 
	of particles to move along with the ether. This would cause 
	an additional drift caused by the overall 
	``ether wind'', that is not present in the Newton-Madelung 
	equation \eqref{eq:Madelung}. 
	\par 
	However, according to the current state of knowledge, by which 
	we mean the 
	point of view imposed by the Einstein equivalence principle and 
	the related non-Euclidean geometry of spacetime 
	(see \cites{Carroll,Wald} 
	for an introduction to general relativity, \cites{Kriele, Sachs} for 
	a more mathematical treatment), a similar argument can 
	be made explaining the noise term \eqref{eq:noise}. That is, if we 
	assume the existence of gravitational waves that are too weak 
	to have a directly observable influence on macroscopic 
	objects, yet strong enough to have an influence on microscopic
	particles such as electrons.
	\par 
	Consider the following, purely relativistic gedankenexperiment: 
	Say we have a physical, inertial 
	observer Alice who perceives her surroundings as 
	having, for instance, a flat geometry%
	\footnote{This means that Alice does not observe any 
	gravitational lensing 
	or deviation from straight-line motion of macroscopic, 
	unaccelerated objects.}
	and who, by some miraculous power, 
	is able to sense the position of an otherwise freely moving 
	particle without disturbing it. Note that this is not a contradiction
	to the Heisenberg inequality, as explained in 
	\thref{Rem:idealdetectors}. 
	If the sufficiently 
	weak gravitational waves are more or less random and there is 
	no gravitational recoil, the particle will
	move geodesically in the actual geometry, but this will not 
	be a straight 
	line according to Alice's perceived, macroscopic geometry. 
	If there is gravitational recoil, the particle
	might not move geodesically and could in principle loose or gain 
	mass depending highly on the relation between the spacetime 
	geometry and the mass of the particle. Either way, Alice would 
	describe the motion of the particle as random and she would have 
	to resort to a statistical description,
	possibly taking the shape of the Madelung equations. 
	Just as in the case of the droplet, the apparently random 
	behavior would 
	be caused by a highly complicated, non-linear underlying dynamics, 
	very susceptible to initial conditions, yet would also be 
	deterministic. Alice, being aware of the underlying physics, 
	would have to construct a model for
	\emph{geometric noise}, that is noise caused by seemingly 
	\emph{random small-scale curvature irregularities in 
	spacetime}. 
	\par  
	While we are aware of the radicality of this ansatz, it 
	appears plausible 
	to us that the Madelung equations and thus also the 
	Schr\"odinger equation 
	could be a model of geometric noise. The fact that a droplet 
	on a vibrating fluid bath is a quantum-mechanical analogue appears 
	to be more than mere coincidence, considering that space and 
	time cannot be 
	assumed to be adequately described by special relativity on the scale 
	of the Bohr radius without severe extrapolation. Even though we do not 
	expect general relativity to be valid at the quantum scale, 
	the thought 
	experiment shows how someone only trained in relativity theory might 
	interpret quantum behavior. Moreover, this conceptual approach 
	can potentially resolve the old question why the electron 
	surrounding a hydrogen nucleus does not 
	radiate, which would cause the atom to be instable \cite{Bohr1}, 
	and why a description employing the Coulomb force works well, 
	despite it only being valid in electrostatics: 
	The electron is standing 
	almost still with respect to the nucleus, but the local 
	spacetime around the nucleus is non-static. In the 
	hydrodynamic analogy, 
	it is like a ball caught in a vortex of a vibrating fluid, which 
	in this case is spacetime itself. The ball does not move much 
	with respect 
	to the fluid, but the fluid does move with respect to an 
	outside observer at rest. 
	\par 
	A geometric origin of the noise term \eqref{eq:noise} has already 
	been proposed by Delphenich \cite{Delphenich}, but, to our 
	knowledge, no satisfactory derivation has been proposed yet. 
	The proposal that quantum behavior is 
	caused by random fluctuations of some microscopic `fluid' goes back 
	to Bohm and Vigier \cite{Bohm3}. In his model of stochastic 
	mechanics, Nelson gave a
	similar interpretation \cite{Nelson1}. Tsekov 
	has formulated his \emph{stochastic interpretation} of the
	Madelung equations as follows: 
	``$[\dots]$  the vacuum  fluctuates permanently  and for  this  reason 
	the trajectory of a particle  in  vacuum is random.  If  the  
	particle  is, however,  too  heavy  the vacuum  fluctuations 
	generate  negligible forces  and  this  particle  obeys  the 
	laws  of  classical  mechanics.'' \cite{Tsekov1}
	Note that the word 'forces' is better replaced by 'deviations from the 
	macroscopic metric' in the interpretation we propose. Ultimately 
	this interpretation should be supported by a mathematical 
	derivation of the Madelung equations from a relativistic model 
	of random irregularities in spacetime curvature. 
	\begin{Question}
		If quantum behavior is caused by random small-scale 
		curvature irregularities in spacetime, how 
		is the noise term to be derived? 
	\end{Question}
	We do not believe that such a derivation, if it exists, is 
	currently within reach and thus caution against any attempts to 
	find it. Even if the hypothesis of quantum behavior being caused 
	by gravitational waves is correct, it appears
	doubtful that the Einstein equation holds on the quantum scale 
	and thus one lacks the basic equations to model the gravitational 
	waves. Even if they are known, one will most likely be faced with 
	a system of non-linear partial
	differential equations for which no general solution can be found 
	and then one would still have to find a way to model the 
	randomness. Clearly superposition of
	waves is only applicable if the differential equation is linear, 
	which makes modeling the randomness a non-trivial task already 
	for Ricci-flat plane waves. 
	Moreover, if one works in the linear approximation, in general one 
	encounters arguably unphysical singularities in the metric 
	\cite{Bondi}. 
	\par 
	Ultimately, a deep question that needs to be addressed in 
	this interpretation of quantum mechanics is how the violation 
	of Bell's inequality is achieved. 
	Ballentine traces the violation of Bell's inequality in 
	quantum mechanics back to the locality postulate used in 
	the derivation of the inequality \cite{Ballentine}*{\S 20.7}.  
	\begin{Postulate}[Bell-Locality]
	\label{Post:Bell}
		If two spatially separated measurement devices $A$ and $B$ 
		respectively measure 
		the observables $a$ and $b$ of an ensemble of two distinct, 
		possibly indistinguishable particles, then the 
		result of $b$ obtained by $B$ does not change as a different 
		observable $a'$ 
		is measured by A and vice versa. 
	\end{Postulate} 
	If we assume that the stochastic interpretation is correct, 
	then it appears to us that there are two possible resolutions 
	to prevent actual so-called ``actions at a distance''. 
	\par
	The first one is that, as in the case of 
	the droplets, the particle itself creates gravitational waves and 
	this in turn 
	influences the motion of other particles, which might appear like a 
	non-local interaction. This approach appears slightly implausible 
	to us, since this could lead to a fluctuation in the mass of 
	the particle, which is not 
	observed. In addition we would naively expect such waves to 
	travel approximately at the speed 
	of light with respect to the macroscopic metric, but 
	\thref{Post:Bell} and thus 
	Bell's argument also includes spacelike separated 
	measurements \cite{Ballentine}*{\S 20.4}. 
	\par 
	The second, to our mind more plausible resolution is to drop 
	an assumption that is implicit in most modern physical 
	theories, namely that a region of space (relative to a 
	physical observer) containing particles is topologically simple 
	on mesoscopic and microscopic scales. 
	The suggestion that there is a connection between topology 
	and entanglement has recently been made by van 
	Raamsdonk \cite{Raamsdonk}, but, to our knowledge, 
	goes back to Wheeler 
	\cites{Wheeler1, Wheeler2}. In that case, we would
	not only have to renounce the statement that spacetime is flat 
	at the quantum 
	scale, but also that it can be adequately modeled by an open subset of 
	$\R^4$. 
	So the idea is that 
	handles in spacetime are observed as entangled 
	particles and the system satisfies both the principle of causality and 
	locality as implemented in the theory of relativity. This 
	necessitates the 
	view of fundamental particles as geometric and topological 
	spacetime solitons, as in Wheeler's 
	``geometrodynamics'' \cites{Wheeler1, 
	Wheeler2}. 
	Then \thref{Post:Bell} is not applicable as the  
	particles are not distinct and 
	thus Bell's inequality can be violated even if \thref{Post:Bell} 
	is true. The non-locality observed for entangled particles is 
	then not real, but only apparent, caused 
 	by interactions of the particles with the measurement apparatus 
 	and a naive conception of space and time.
 	\par 
 	However, in order to overcome the speculative nature of 
 	this discussion, we suggest that the proper implementation of spin 
 	and the treatment of multi-particle systems 
 	in the Madelung picture is carried out first. Following the 
 	discussion in section \ref{sec:Newton}, this might require a 
 	detour through the relativistic theory and the Newtonian limit.  
 	
\section{Modification: Particle Creation and Annihilation}
\label{sec:annihilation}

As stated in the introduction, the Madelung equations can be naturally 
modified to study a wider class of possible quantum systems. For 
instance, one can consider rotational forces and `higher order quantum 
effects' by viewing the noise term \eqref{eq:noise} as the first order in 
a Taylor approximation in $1/m$ around $0$ of a non-linear operator in 
$\rho$ and its derivatives. The modification we propose here is 
of conceptual nature and intended to be applied in the generalization 
of the formalism to many particle systems. Though we do not wish to 
fully address this generalization here, we remark that, due to 
the symmetrization postulate \cite{Ballentine}*{\S 17.3}, the concept 
of spin needs to be properly implemented in the Madelung picture first, 
to be able to study systems with multiple mutually indistinguishable 
particles. The results obtained in the linear operator formalism can 
serve as a guide (see \cite{Ballentine}*{\S 18.4}), but should also 
be questioned. 
\par
The phenomenon of particle creation and annihilation is not one 
that requires a relativistic treatment per se 
\cite{Ballentine}*{\S 17.4}, despite the fact that it is 
most commonly 
considered within relativistic quantum theory. Besides, the treatment 
in quantum field theory is also not free of problems (see e.g. 
\cite{Nicolic}*{\S 9.5}). 
This raises the question how this phenomenon should be modeled in 
the Madelung picture. Following our discussion in the previous section 
on page \pageref{sec:interpretation}, 
it becomes obvious that the continuity equation
\begin{equation}
	\partd{\rho}{t} + \nabla \cdot \left( \rho \vec X \right) = 0 \, ,
	\label{eq:continuity1}
\end{equation}
needs to be modified, as, once normalized, it leads to the conservation of probability 
\begin{equation}
	\int_{\Omega_t} \iota^*_t \rho \,  \d^3 x = 1 \quad \forall t \in \R 
	\, .
\end{equation}
In fluid mechanics \eqref{eq:continuity1} 
is the conservation of mass \cite{Chorin}*{\S 1.1}. 
To model a change in mass of the fluid, e.g. due to chemical 
reactions, one includes a source term 
\begin{equation}
	\tilde u = 
	\partd{\rho}{t} + \nabla \cdot \left( \rho \vec X \right) \, ,
	\label{eq:u} 
\end{equation} 
which implies that
\begin{equation}
	\partd{}{t} \int_{\Omega_t} \iota_t^*\rho \, \d^3 x = 
	\int_{\Omega_t} \iota_t^* \tilde u  \, \d^3 x \, ,
	\label{eq:source}
\end{equation} 
by the Reynold's transport theorem (modulo questions of convergence). 
In quantum mechanics, equation \eqref{eq:source} can be interpreted 
as stating that the probability of finding the 
particle anywhere changes with time, which is the desired modification 
to the continuity equation. More precisely, $\tilde u$ should be 
replaced by a smooth, possibly trivial operator $u$ 
applied to $\rho$ and $X$, in the sense that 
\begin{equation}
	u \left( \rho, X \right) \colon \mathcal Q \to \R
\end{equation}
is smooth for all smooth $\rho$ and $X$. That the domain of 
$u \left( \rho, X \right)$ is  $\mathcal Q $, rather than, e.g. 
$ \mathcal Q \cross  \mathcal Q $, is required by the principle 
of locality. Moreover, since probabilities are nonnegative and not 
greater than $1$, we also have to demand
\begin{equation}
	\int_{\Omega_t} \iota_t^*\rho  \, \d^3 x \in [0,1] \subset 
	\R \quad \forall t \in I \, . 
\end{equation} 
Thus the Madelung equations for one Schr\"odinger particle that can 
be created and annihilated (e.g. by formation from or disintegration 
into gravitational waves, see section \ref{sec:interpretation}) consist 
of the Newton-Madelung equation \eqref{eq:Madelung}, 
the irrotationality of the drift field \eqref{eq:irrotX} and the modified 
continuity equation 
\begin{equation}
	\partd{\rho}{t} + \nabla \cdot \left( \rho \vec X \right) = u 
	\left( \rho, X\right)
	\, ,
	\label{eq:newu} 
\end{equation}
where the precise form of $u$ is still unknown. Due to the scaling 
invariance of the Newton-Madelung equation, this modification does 
not change the underlying dynamics.
We also remark that the requirement for the equations to separate 
for isolated ensembles puts restrictions on $u$ (see page 
\pageref{eq:noise}sqq.).   
\begin{Proposition}
	\label{Prop:annihilation}
\begin{subequations}
	Let $\left( \mathcal Q , \d \tau, \delta, 
	\mathcal O \right)$ be a Newtonian spacetime with 
	$b_1 \left(\Omega _t \right)= 0$ for all $t
	\in I$, as defined in \eqref{eq:Omega}, let $X \in \mathfrak X _{Nt} 
	\left( \mathcal Q \right)$ be a Newtonian observer vector field, 
	$\vec F \in \mathfrak X _{Ns} \left( \mathcal Q \right)$ be 
	a Newtonian spacelike vector field, $\rho \in C^\infty 
	\left( \mathcal Q , \R_+ \right)$ a strictly positive, real 
	function,
	\begin{equation}
		u \colon C^\infty \left(\mathcal Q , \R_+ \right) \cross 
		\mathfrak X_{Nt}\left( \mathcal Q\right) 
		\to C^\infty \left(\mathcal Q , \R \right) 
		\colon \left(\rho, X \right) 
		\to u \left( \rho, X \right)
		\label{eq:opu}
	\end{equation}
	an operator and $m, \hbar \in \R_+$. \\
	Then the irrotationality of 
	$X$ \eqref{eq:irrotX} and $F$ \eqref{eq:irrotF} 
	imply that $\exists \varphi, V \in C^\infty \left( \mathcal Q, 
	\R \right)$
	such that equations \eqref{eq:gradvarphi}, \eqref{eq:V} on page 
	\pageref{eq:gradvarphi} hold and by setting 
	$\Psi := \sqrt \rho \, e^{-\i \varphi}$, 
	\begin{equation} 
		\xi \left( \Psi \right) 
		:= \frac{\hbar}{2 \abs{\Psi}^2} \, 
		u \left( \abs{\Psi}^2, \partd{}{t} + \frac{\hbar}{2 \i m}
		 \left( \frac{\nabla \Psi}{\Psi}
		- \frac{\nabla \Psi^*}{\Psi^*} \right) \right) \, , 
		\label{eq:xi}
	\end{equation}
	the Newton-Madelung equation \eqref{eq:Madelung} together with 
	equation \eqref{eq:newu} imply  
	\begin{equation}
		\i \hbar \partd{\Psi}{t} = 
		- \frac{\hbar^2}{2m} \Delta \Psi + V \Psi + \i \xi 
		\left( \Psi \right)
		\Psi 
		\, . 
		\label{eq:SExi}
	\end{equation}
	Conversely, for 
	$\Psi \in C^\infty\left( \mathcal Q, \C \setminus \lbrace 0 \rbrace
	\right)$, $V 
	 \in C^\infty\left( \mathcal Q, \R 
	\right)$ and 
	\begin{equation}
		\xi \colon 
		C^\infty\left( \mathcal Q, \C \setminus \lbrace 0 \rbrace \right)
		\to C^\infty\left( \mathcal Q, \R \right)
		\colon  \Psi \to 
		\xi \left( \Psi \right)
	\end{equation}
	satisfying \eqref{eq:SExi}, define 
	$\rho := \abs{\Psi}^2$, $\vec F$ via \eqref{eq:V}, $u$ via 
	\eqref{eq:opu} as well as \eqref{eq:xi} 
	and $\vec X$ via \eqref{eq:XfromPsi} such that  
	$X:= \partial/ \partial t + \vec X 
	\in \mathfrak X _{Nt} \left( \mathcal Q \right)$. 
	Then \eqref{eq:Madelung}, \eqref{eq:newu}, 
	\eqref{eq:irrotX} and \eqref{eq:irrotF} hold.  
\end{subequations}
\end{Proposition}
\begin{Proof}
\begin{subequations}
	The proof is entirely analogous to the one of \thref{Thm:equivalence} 
	on page \pageref{Thm:equivalence}. Instead of \eqref{eq:ImSE}, we get
	\begin{equation}
		2 R \partd{R}{t} - \frac{\hbar}{m} \left(2 R \nabla R \cdot 
		\nabla \varphi + R^2 \Delta \varphi  \right) = u \left( 
		R^2, \partd{}{t} - \frac{\hbar}{m} \nabla \varphi \right) \, .
	\end{equation}
	Using \eqref{eq:XfromPsi} on page \pageref{eq:XfromPsi},  
	definition \eqref{eq:xi} above and formula 
	\eqref{eq:DeltaPsi} for $\Delta \Psi$ on page \pageref{eq:DeltaPsi}, 
	we obtain
	\begin{equation}
		\hbar \partd{R}{t} = - \frac{\hbar^2}{2m} \Im \left( e^{\i \varphi 
		} \Delta \Psi \right) + R \xi \left( \Psi\right) \, .
	\end{equation}
	Together with the real part of $e^{\i \varphi } \Delta \Psi $ 
	\eqref{eq:ReSE} we indeed get \eqref{eq:SExi}. The reverse 
	implication is also proven in full analogy to 
	\thref{Thm:equivalence}. 
\end{subequations}
\end{Proof}
If we now define an operator $\hat \Xi $ acting on $\mathcal S 
\left( \R^3, \C \setminus \lbrace 0 \rbrace 
\right)$ via 
\begin{equation}
	\hat \Xi \Psi_t := \left( - \frac{\hbar^2}{2m} \Delta  + V 
	+ \i \xi \left( \Psi_t \right) \right) \Psi_t
\end{equation}
for $\Psi_t \in \mathcal S \left( \R^3, \C \setminus \lbrace 0 \rbrace 
\right)$, then $\hat \Xi$ is usually non-linear and need not even 
be defined on $\mathcal S \left( \R^3, \C\right)$. 
The Schr\"odinger equation modeling particle
creation and annihilation 
\begin{equation}
	\hat E \Psi_t = \hat \Xi \Psi_t
\end{equation}
can then not be recast into an eigenvalue equation for $\hat \Xi$, 
as the separation ansatz will not work. We have thus proposed a 
physically reasonable model in which
the current axiomatic framework of quantum mechanics breaks down 
(see section \ref{sec:operator} on page \pageref{sec:operator} and 
\cite{Ballentine}*{\S 2.1}). 
	
\section{Conclusion}
\label{sec:conclusion}

In the introductory discussion we have argued that the use of a 
quantization algorithm in the formulation of quantum mechanics is a 
strong indication that quantum mechanics and thus quantum theory as a 
whole is, as of today, an incomplete theory. We also suggested that 
the identification of fundamental geometric quantities is a promising 
path to overcome this somewhat unsettling feature, as these quantities 
will inevitably be part of a new axiomatic framework for the theory. 
We then proceeded in section \ref{sec:Newton} by constructing a 
Newtonian spacetime on which we then formulated the Madelung equations 
in section \ref{sec:equivalence}. This construction enabled us to proof 
a local equivalence of the Madelung equations and the 
Schr\"odinger equation for irrotational forces. 
By relating the Madelung equations to the 
linear operator formalism thereafter, we showed that the Madelung 
equations naturally explain why the position, momentum, energy and 
angular momentum operators take the shape commonly found in quantum 
mechanics textbooks. These results strongly indicate that the Madelung 
equations formulated on a Newtonian spacetime provide the natural 
mathematical basis for quantum mechanics and that this basis should 
include the relevant aspects of Kolmogorovian probability theory. In 
section \ref{sec:interpretation1} we gave a formal discussion of the 
Madelung equations that can be used for practically interpreting and 
applying the formalism, as well as extending the mathematical model. 
We then proceeded in section \ref{sec:interpretation2} by speculating 
that quantum mechanics provides a statistical model for spacetime 
geometric noise, which is a variant of the stochastic interpretation 
developed by Bohm, Vigier and Tsekov. To give an example how to naturally 
extend the Madelung equations, we proposed an unfinished model for 
particle creation and annihilation for single-Schr\"odinger particle 
systems in section \ref{sec:annihilation}. We observed that this can 
lead to a non-linearity in the resulting Schr\"odinger equation and 
thus makes the linear operator formalism inapplicable. 
\par 
Some of our 
results have been summarized in the table below. The abbreviations 
QM and GQT stand for quantum mechanics and geometric quantum theory, 
respectively. 
\begin{longtable}[h]{M{0.185 \textwidth} M{0.32 \textwidth} 
						M{0.32 \textwidth}M{0.055 \textwidth}}   
	\hline
	Subject & 
	Textbook/Copenhagen QM & 
	GQT in Newtonian limit for Schr\"odinger particles & 
	cf.\\
	\hline \hline 
	Spacetime model &
	Newtonian spacetime   
	$( \mathcal Q, \d \tau, \delta, \mathcal O )$ 
	with $\Omega_t = \R^3$ (modulo sets of measure zero)
	$\forall t \in I$, but implicit
	&
	Newtonian spacetime 
	$\left( \mathcal Q, \d \tau, \delta, \mathcal O \right)$ 
	& \S \ref{sec:operator} \& \S \ref{sec:Newton} 
	\\
	\hline 
	Single particle state & 
	(spinor) wave function $\Psi$ & 
	probability density $\rho$ and drift field $X$ & 
	\S \ref{sec:equivalence} \& \S \ref{sec:interpretation1} 
	\\  
	\hline
	Probability theory used	& 
	von Neumann with projection postulate & 
	Kolmogorov with measure $\int \d^3 x \, \iota_t^* \rho$ 
	applied on Borel sets or Lebesque sets of $\Omega_t \subseteq \R^3$ & 
	\S \ref{sec:operator} \, \& 
	Post. \ref{Post:Kolmogorov} \\ 
	\hline  
	Observables &	inner products of wave functions, 
	elements in the spectrum of (linear) 
	endomorphisms of a Hilbert space $\mathcal H$ & probabilities and 
	expectation values of real-valued functions on $\mathcal Q$ 
	(possibly depending on states) & 
	as above
	\\ 
	\hline  
	Measurement problem	&	
	unresolved; in Copenhagen interpretation measurement causes 
	`wave function collapse' &	
	not an issue; wave function 
	is a mathematical tool encoding information on ensembles of 
	particles; measurement itself is not modeled & 
	\S \ref{sec:interpretation1} \& 
	Rem. \ref{Rem:idealdetectors}
	\\
	\hline 
	Wave-particle duality	& particle identified with 
	wave function $\Psi$; interpreted as actual wave
	in Copenhagen interpretation & 
	makes no statement on internal structure of particles; 
	treats them as effectively point-like & 
	Rem. \ref{Rem:sparticles}
	\\
	\hline
	Superposition &	fundamental principle of QM; 
	implemented via linear operator formalism on some Hilbert space
	&	
	not a principle; only 
	sensible, 
	if $\Psi$ exists and dynamical evolution equation 
	in $\Psi$ is linear & 
	\S \ref{sec:operator}, \S \ref{sec:interpretation1} \& \S 
	\ref{sec:annihilation}   
	\\
	\hline
	Classical correspondence & 
	QM supposedly yields Newtonian mechanics in the limit $\hbar \to 0$ & 
	a Newtonian probability theory is obtained in large mass 
	approximation & 
	Equs. \eqref{eq:pNewton} \& 
	Post. \ref{Post:nq-limit}
	\\
	\hline
	Canonical
	quantization &	
	ill-defined scheme to obtain dynamical equations 
	from Newtonian mechanics
	&	
	rejected; instead
	dynamical equations are postulated, justified by   
	arguments and empirical evidence
	& 
	\S \ref{sec:intro} \& 
	\S \ref{sec:equivalence}   
	\\
	\hline
	Uncertainty	relation & 
	interpreted as fundamental uncertainty in measurable 
	position and momentum & 
	relation formally derivable, but interpretation not supported; 
	no restriction on maximal precision of measurement 
	on theoretical level & 	
	Rem. \ref{Rem:idealdetectors} \& \S \ref{sec:interpretation1} \\
	\hline
	Particle creation \& annihilation & not possible; in QFT
	via $2$\textsuperscript{nd} quantization formalism & possible 
	in principle
	via modification of continuity equation & 
	\S \ref{sec:annihilation} \\
	\hline
	Fundamental Theory? & yes, in Newtonian limit & 
	no, phenomenological & \S \ref{sec:intro} \&
	\S \ref{sec:interpretation2} \\
	\hline
\end{longtable}   
\noindent
Despite all of these remarkable successes of the Madelung picture, 
there are still many open problems that need to be addressed to 
complete it and put quantum theory on a new foundation. From a 
mathematical point of view, the most important one is formulated by 
\thref{Que:existence} on page \pageref{Que:existence}. We are currently 
working on the proper generalization of the Madelung equations to 
the relativistic setting, which is of conceptual importance due to the 
principle of relativity as discussed in the beginning of section 
\ref{sec:Newton}. However, there are many potentially fruitful paths 
of extending the Madelung equations in the non-relativistic setting 
already. How is spin to be geometrically implemented? How does 
the generalization to many particle systems work? 
How exactly do we model particle creation and annihilation? 
Finally, there remains the question of interpreting the Madelung 
equations: How does the hydrodynamical quantum analogue discovered by
Couder et al \cites{Couder1,Couder2,Couder3,Eddi,Fort,Protiere} relate 
to the actual behavior of quanta? How is matter related to spacetime 
geometry on the quantum scale? 
\par 
To answer these questions, the non-quantum limit, the existing 
literature on quantum theory formulated in the linear operator formalism 
(e.g. \cites{Weinberg0,Weinberg1,Ballentine,Hestenes1})
as well as already existent results obtained in Bohmian mechanics 
(e.g. \cites{Bohm0,Bohm1,Bohm2,Bohm3,Bohm4,Bohm5,Bohm6,Holland}) 
will be of use.

\section*{Acknowledgements}   

I would like to thank the following people who have made 
this work possible in various ways: Ina, Heiko and Marcel 
Reddiger, Helmut and Erika Winter, Erich and Rita Reddiger, 
Katie, Timothy and Camden Pankratz, Whitney Janzen-Pankratz, 
Viktor Befort, Viola Elsenhans, Adam Murray, Marcus Bugner, 
Alexander Gietelink Oldenziel, Christof Tinnes, Knut Schn\"urpel, 
Thomas K\"uhn, Arwed Schiller, Gerd Rudolph, Dennis Dieks, Gleb 
Arutyunov, Wolfgang Hasse and Maaneli Derakhshani. Moreover, I am 
grateful to Adam Murray and Markus Fenske for their help in 
correcting the manuscript and the two anonymous referees for 
their constructive comments. Thanks goes also to the dear 
fellow who wrote the Wikipedia article on the Madelung equations, 
without which I might have not stumbled upon them and realized 
their importance. 

\addsec{References}

\end{document}